\documentclass[11pt,a4paper]{article}
\usepackage{jheppub}
\usepackage[english]{babel}	          
\usepackage{hyperref}

\newcommand{\beq}{\begin{equation}}
\newcommand{\eeq}{\end{equation}}
\newcommand{\BC}{\begin{center}}
\newcommand{\EC}{\end{center}}
\newcommand{\BF}{\begin{figure}}
\newcommand{\EF}{\end{figure}}
\newcommand{\bs}{\bigskip}
\newcommand{\nn}{\nonumber}
\newcommand{\dd}[1]{\ensuremath{\mathrm{d}{#1}}}
\newcommand{\ud}{\mathrm{d}}
\def\bea{\begin{eqnarray}}
\def\eea{\end{eqnarray}}
\def\bean{\begin{eqnarray*}}
\def\eean{\end{eqnarray*}}

\title{Lovelock solutions in the presence of matter sources}

\author[a]{Yannis Bardoux,}
\author[a,b]{Christos Charmousis,}
\author[c]{Theodoros Kolyvaris}

\affiliation[a]{\href{http://www.th.u-psud.fr/}Laboratoire de Physique Th\'eorique
CNRS UMR 8627, Universit\'e Paris-Sud 11 91405 Orsay Cedex, France}
\affiliation[b]{\href{http://www.lmpt.univ-tours.fr/}Laboratoire de Math\'ematiques et Physique Th\'eorique (LMPT) CNRS  UMR 6083 Universit\'e Fran\c cois Rabelais - Tours, France}
\affiliation[c]{\href{http://www.physics.ntua.gr/index_en.html}Department of Physics, National Technical University of Athens,
Zografou Campus GR 157 73, Athens, Greece}
\emailAdd{yannis.bardoux@th.u-psud.fr}
\emailAdd{christos.charmousis@th.u-psud.fr}
\emailAdd{teokolyv@central.ntua.gr}

\date{today}
\abstract{For a large class of space and time-dependent warped geometries we 
find the general solution of the 6-dimensional Einstein-Gauss-Bonnet equations in the presence of $p$-form matter fields. This is done under two conditions on the matter sector which we show impose the integrability of the full system.  Solutions are classified and known black hole limits are found. It is shown that Lovelock gravity restricts drastically the possible horizon geometries and allowed matter sources. In fact, we show that if we allow only for solutions of asymptotically flat falloff behaviour, and no fine-tuning of coupling constants,  then the only permissible black hole is that of Boulware-Deser with electromagnetic charge. The situation of 6 dimensional Lovelock gravity is therefore almost identical to 4 dimensional General Relativity. The gravitational horizon constraints lead us to find static solutions involving 3-form matter fields in anti de Sitter space which are also new to General Relativity along with other cosmological and black string type of solutions.}

\begin{document}
\maketitle
\flushbottom

\section{Introduction}

Uniqueness theorems, such as Birkhoff's theorem are at the heart of physical applications in General Relativity (GR). This theorem establishes, for spherical symmetry and in the vacuum, the uniqueness of the Schwarzschild black hole which means that  the gravitational field of a source of spherical symmetry is that of the Schwarzschild metric, be it a very extreme object as that of a black hole or a common star like the sun. This makes sense because the Schwarzschild radius of the sun is of the order of 3 km whereas its actual star radius could fit twice the distance inbetween the Earth and the moon. Thus solar system experiments where GR is put to the test rely on this unique metric field and GR succeds local gravity tests with flying colours \cite{Will}. Writting up Schwarzschild in isotropic coordinates and expanding the metric components gives with ease the leading Eddington parameters in the parametrised post-newtonnian (PPN) approximation of the theory. This unique situation is unlike other gravity theories like Brans-Dicke for example \cite{Brans}, \cite{Damour}, where black holes and stars do not have share the same gravitational field. This is due to the breakdown of Birkhoff's theorem and results in differing Eddington parameters with GR and thus  leads to tension between experiment and theory \cite{Gilles}. This uniqueness property of GR is due to the presence of only spin-2 massless excitations and the absence of scalar excitations which would permit spherically symmetric breather mode fluctuations to be excited. The uniqueness of spin-2 excitations is closely related to another theorem of uniqueness in GR, Lovelock's theorem. This theorem states that in 4 dimensions the only metric theory action (of second order in the metric), endowed with a Levi-Civita connection and yielding second order field equations and Bianchi identities is GR with cosmological constant \cite{lovelock}. At the perturbative level the theorem results in the presence of only 2 spin-2 graviton fluctuations and in particular in the absence of the troublesome conformal mode which is not excited. No other such pure spin 2 theories are known with this property although the theorem does not completely rule out the possibility.

Lovelock's theorem in higher dimensions \cite{lovelock}, \cite{Zumino} gives the higher dimensional version of GR  in 4 dimensions. In order to allow for the most general second order equations of motion specific Lovelock densities have to be added to the gravitational action, one for each 2 extra dimensions, hence the Gauss-Bonnet term for 5 or 6 dimensional spacetime (for a review see \cite{Deruelle}, \cite{charm}, \cite{Garraffo}). Lovelock theory has found important implications in string theory \cite{Zwiebach} as the leading order quantum gravity correction \cite{Metsaev}, \cite{Gross} and also in the realm of braneworlds where its presence can provide richer GR phenomenology (see for example \cite{Sasaki}, \cite{Z1}, \cite{Kofinas}). A natural question arises: Since Lovelock theory seems closely related to Birkhoff's theorem in GR does this theorem also hold in higher dimensions? The answer is positive as was first shown by Wiltshire in the early 80's \cite{wilt1}. A minimal generalisation and the application to braneworld cosmology was later obtained \cite{fax1} with the result generalised for Lovelock's theory in \cite{zegers}. Several generalisations and methods were pursued \cite{deser1} but let us step back and question the physical application of this theorem in higher dimensions. In 5 dimensional brane cosmology the theorem states that a distributional braneworld endowed with perfect fluid matter has a uniquely determined trajectory \cite{bowcock}. As such it is found that the solution involves a time dependent hypersurface which evolves in a static black hole background. Therefore, although one would expect at the appearence of an extra dimension, a novel degree of freedom to appear (the radion)-the theorem confirms the opposite. This timelike trajectory  is exactly the Hubble expansion factor as is the situation for 4-dimensional General Relativity. This explains the presence of the modified FRW equations and the absence of the radion mode for braneworld cosmology. The background is again a higher dimensional static adS Schwarzschild black hole solution \cite{bowcock} (and the Boulware-Deser-Cai black hole \cite{Boulware}, \cite{Cai}  for the Lovelock version \cite{fax1}). In 6 dimensions the elliptic version\footnote{This elliptic statement will become clearer in the next section} of this theorem yields the unique flat, de Sitter and anti de Sitter codimension-2 vacua of the theory \cite{papa1}.

In higher dimensions Birkhoff's theorem can also be generalised with respect to the permitted horizon geometries. In fact, although in 4 dimensions the only 2-dimensional horizon sections are of maximal symmetry, in 6 dimensions and higher these spaces can be generic Einstein spaces. It was shown by Gibbons and Hartnoll \cite{gib} that {\it any Einstein space} is permissive as a horizon (with certain conditions on the curvature scales). For example one can have,
\beq
\label{euclidean}
ds^2=-V(r)\ud t^2+\frac{\ud r^2}{V(r)}+r^2 \left(f(\rho)\ud\tau^2+\frac{\ud\rho^2}{f(\rho)}+\rho^2 \ud\Omega^2_{II} \right)
\eeq
with $f(\rho)=1-\frac{\mu}{\rho}$ an euclidean Schwarzschild metric as horizon or again
\beq
ds^2=-V(r)dt^2+\frac{dr^2}{V(r)}+r^2 dT^2_{IV}
\eeq
with a horizon simply flat toroidal geometry. In both metrics the black hole horizon {\it is the same},
$V(r)=-\frac{\Lambda}{10} r^2-\frac{m}{r^3}$. The black hole is not influenced by the geometry details of the horizon. Other examples of such metrics are black string metrics where we double Wick rotate (\ref{euclidean}) or simply add $N$ flat dimensions to a 4 dimensional Schwarzschild metric (they are solutions of an elliptic rather than hyperbolic problem),
\beq
\label{euclidean2}
ds^2=\sum_N \ud z_N^2+ \left(-f(\rho) \ud\tau^2+\frac{\ud\rho^2}{f(\rho)}+\rho^2 \ud\Omega^2_{II} \right) \ .
\eeq
This large degeneracy although useful at first in producing multitude of solutions in differing contexts often hides inconsistency as was first shown by Gregory and Laflamme in their celebrated black string instability \cite{GL} (see also the general analysis in \cite{gib} for the hyperbolic problem). One may question the situation in Lovelock theory? Is this horizon degeneracy due to the absence of the additional terms in the action in more than 4 spacetime dimensions? This question was answered recently in two papers firstly establishing the spherically symmetric black hole solution \cite{Dotti} and then in \cite{Bogdanos:2009pc} where the general solution (without assuming staticity) was found, the horizon structure was given  and specific examples analysed. The most important result of this study is that the large degeneracy of permitted horizon metrics of higher dimensional general relativity is lifted once the full general action of Lovelock is taken into account. The novel black holes not only have horizons which are Einstein metrics but these, when different from the maximally symmetric solutions modify the black hole potential and the black hole asymptotics and impose a particular relation for the 4 dimensional Weyl tensor.

In the present paper we will ask the following question for 6 spacetime dimensions{\footnote{Here we emphasize that it is essential for the horizon geometry to be at least 4 dimensional in order for 4 dimensional curvature quantities such as the Weyl tensor to be non-trivial}}: how does the presence of matter  modify the known solutions of Lovelock theory? Based on the integrability conditions found in \cite{Bogdanos:2009pc} and imposing them anew we will find the general solution involving scalar fields, a $U(1)$ EM field or 3 forms (higher forms can be obtained by a common duality relation). Some of the solutions we will find will be novel even in the GR limit. These can be black holes involving 3-forms and scalar fields living on the black hole horizon geometry.

Our matter action will involve exact $p$-forms $\mathcal{F}$ as $\int_\mathcal{M} \mathcal{F} \wedge \star \mathcal{F}$ for which there exists a potential  $\mathcal{A} \in \Lambda^{p-1}\left(\mathcal{M}\right) $ such that $\mathcal{F} = d \mathcal{A}$ where $\Lambda^p\left(\mathcal{M}\right)$ is the space of $p$ forms of $\mathcal{M}$. Given that we restrict our analysis to $D=6$ dimensions, $p=1,2,..,6$, $p=1$ corresponds to a kinetic term for a scalar field played by the potential $\mathcal{A}$ and $p=2$ is the usual electromagnetic interaction. $p=3$ corresponds to self-dual 3-forms. The higher $p$-forms will be related to the lower order ones via a duality transformation (this includes the cosmological constant which will be included in the action).

In the next section we give our hypotheses and set up our field equations and a duality symmetry. We then give the 3 classes of solutions in a generic form, obtaining the relevant potentials and geometric conditions on the horizon sections. In section 4 we construct specific examples and finally we conclude.

\section{General set-up}

\noindent
Consider the Lovelock action in six dimensional spacetime in the presence of an exact gravitating $p$-form, $\mathcal{F} = \frac{1}{p!}F_{A_1 ... A_p} \mathrm{d}y^{A_1} \wedge ... \wedge \mathrm{d}y^{A_p}$, given by

\beq S^{(6)} = \frac{M^{(6)^4}}{2} \int_\mathcal{M} \mathrm{d}^6 x \sqrt{-g^{(6)}} \left[ R - 2\Lambda + \alpha \hat{G} - \frac{\kappa}{p!} F_{A_1 ... A_p}F^{A_1 ... A_p} \right]  \label{action} \eeq
where $M^{(6)}$ is the fundamental mass scale in six-dimensional spacetime, $\kappa$ is the matter coupling constant, $\hat{G}$ the third Lovelock density which is usually dubbed Gauss-Bonnet term,
\beq \hat{G} = R_{ABCD}R^{ABCD} - 4R_{AB}R^{AB} + R^2 \ . \eeq
Uppercase indices refer to six-dimensional coordinates.
With these conventions we vary the action with respect to the metric to derive the Lovelock field equations
\beq \mathcal{E}_{AB} = G_{AB} + \Lambda g_{AB} - \alpha H_{AB} = \kappa T_{AB} \label{EOM1} \eeq
with
\beq \label{matter} T_{AB} = \frac{1}{(p-1)!} F_{A C_1 ... C_{p-1}} F_B^{\ C_1 ... C_{p-1}} - \frac{1}{2p!}g_{AB} F_{C_1 ... C_p}F^{C_1 ... C_p} \eeq
and $G_{AB}$ the Einstein tensor. We have also introduced the Lanczos \cite{Lanczos} or Lovelock tensor \cite{Lovelock2} 
\beq H_{AB} = \frac{1}{2} g_{AB} \hat{G} - 2 R R_{AB} + 4 R_{AC} R^C_{\ B} + 4 R_{CD} R_{\ A \ B}^{C \ D} - 2 R_{ACDE} R_B^{\ CDE} \eeq
which comes from variation of $\hat{G}$ in (\ref{action}).
On the other hand, variation with respect to the potential $\mathcal{A}$ defined as $\mathcal{F}=d\mathcal{A}$ gives $\delta\mathcal{F} = 0$, ie
\beq \partial_{A_p}\left(\sqrt{-g^{(6)}} F^{A_1 ... A_{p}} \right) = 0 \ . \label{EOM2} \eeq

In order to proceed we are now going to choose an appropriate symmetry for the metric and the matter fields. We distinguish between the transverse 2-space, which carries a timelike coordinate $t$ and a radial coordinate $r$, and the internal 4-dimensional space sections which we call $\mathcal{H}$, representing, for example, the horizon geometry in the case of six-dimensional black holes. We will assume that the internal space is endowed with $h_{\mu\nu}(x)$, an arbitrary metric of coordinates $x^\mu$, $\mu = 0,1,2,3$. We will be imposing that the internal space $\mathcal{H}$ and transverse space are locally orthogonal to each other. This is an additional hypothesis with respect to ordinary GR, since $h_{\mu\nu}$ is {\it not} necessarily a homogeneous metric and because our six-dimensional space is \textit{not} necessarily an Einstein space (in GR such an orthogonal foliation is possible for an Einstein metric \cite{gib}). At a loss of a better name we will call this a warped metric Ansatz. Finally, guided by the analogous procedure of analyzing Birkhoff's theorem \cite{bowcock}, \cite{fax1}, we write the metric as
\beq 
\label{metric0}
\mathrm{d}s^2 = e^{2\nu(t,z)} B(t,z)^{-3/4} \left( -\mathrm{d}t^2 + \mathrm{d}z^2 \right) + B(t,z)^{1/2} h_{\mu\nu}(x) \mathrm{d}x^\mu \mathrm{d}x^\nu \ . \eeq
Lowercase greek indices correspond to internal coordinates of the 4-space $\mathcal{H}$. The above metric (\ref{metric0}) encompasses all the above requirements and fixes the symmetries of our spacetime. We then switch the coordinates of the transverse 2-space to light-cone coordinates,
\beq u=\frac{t-z}{\sqrt{2}} \text{ and } v=\frac{t+z}{\sqrt{2}} \eeq
in terms of which the metric reads,
\beq \label{metric} \mathrm{d}s^2 = - 2 e^{2\nu(u,v)} B(u,v)^{-3/4} \mathrm{d}u\mathrm{d}v  + B(u,v)^{1/2} h_{\mu\nu}(x) \mathrm{d}x^\mu \mathrm{d}x^\nu  \ . \eeq

In fact for future reference let us double Wick rotate time coordinate $t=i\theta$ and (admitting it possible) one of the $x^\mu$ coordinates into a timelike coordinate,
\beq 
\label{bstrings}
\mathrm{d}s^2 = e^{2\nu(\theta,z)} B(\theta,z)^{-3/4} \left(\mathrm{d}\theta^2 + \mathrm{d}z^2 \right) + B(\theta,z)^{1/2} h_{\mu\nu}(x) \mathrm{d}x^\mu \mathrm{d}x^\nu 
\eeq
where it is implicitely understood that $h_{\mu\nu}$ is now of lorentzian signature. 
Solutions to the above metric Anzatz can describe, for certain initial conditions, codimension-2 warped black string metrics \cite{papa1} or warped Kaluza-Klein spaces, where the warp factor is precisely $B(\theta,z)^{1/2}$. Furthermore using complex conjugate coordinates we can go to the analogue $(u=\frac{-\theta+i z}{\sqrt{2}}$, $v=\frac{\theta+i z}{\sqrt{2}})$ frame above (\ref{metric}). Loosely speaking the field equations adjucent to the latter (\ref{bstrings}) is the elliptic version of the former hyperbolic problem in (\ref{metric0}). Here, we will give the resolution for the time dependent hyperbolic problem but it is understood that resolution of the elliptic problem follows identically modulo differing boundary conditions \cite{ruthtony}. Example solutions of both metrics (\ref{metric0}), (\ref{bstrings}) will be considered.  

Using the above prescription (\ref{metric}), we now write down the $uu$ and $vv$ left hand side of \eqref{EOM1}
\beq \mathcal{E}_{uu}=\frac{2 \nu_{,u} B_{,u}- B_{,uu}}{B} \left[ 1+\alpha \left( B^{-1/2} R^{(4)}+\frac{3}{2} e^{-2\nu} B^{-5/4} B_{,u} B_{,v} \right) \right]\label{uu} \ ,\eeq
\beq \mathcal{E}_{vv}=\frac{2 \nu_{,v} B_{,v}- B_{,vv}}{B} \left[ 1+\alpha \left( B^{-1/2} R^{(4)}+\frac{3}{2} e^{-2\nu} B^{-5/4} B_{,u} B_{,v} \right) \right]\label{vv} \ .\eeq
As we shall see the factorisable form of these equations is capital for the integrability of the problem and the obtention of exact solutions. In the absence of matter fields it was shown \cite{Bogdanos:2009pc} that three and only three classes of solutions were possible upon choosing $B$ constant, or eliminating one of the two factors in (\ref{uu}) or (\ref{vv}). Our second working hypothesis here will be to preserve $\mathcal{E}_{uu}=0$ and $\mathcal{E}_{vv}=0$ {\it even} when matter-sources are present. For the electromagnetic field $p=2$, this is quite natural and a version of Birkhoff's theorem can be obtained. What however happens for a generic $p$-form? In other words if we impose $T_{uu}=0$ and $T_{vv}=0$ for (\ref{metric}) do we obtain a non-trivial yet reasonable and workable hypothesis?

From (\ref{matter}) taking $T_{uu}=0$ gives,
\beq h^{\rho_1 \sigma_1} ... h^{\rho_{p-1} \sigma_{p-1}} F_{u \rho_1 ... \rho_{p-1}}  F_{u \sigma_1 ... \sigma_{p-1}} = 0  \ (u \leftrightarrow v) \ . \eeq
Since $h_{\rho\sigma}$ is a riemannian metric we obtain
\beq F_{u \sigma_1 ... \sigma_{p-1}} = 0 = F_{v \sigma_1 ... \sigma_{p-1}} \ . \label{IC} \eeq
In particular, for free scalar field, $p=1$, this notation means for example that $F_u = F_v=0$. In other words, we must consider a scalar field, $\mathcal{A}=\phi$, which depends only on the internal coordinates $x^\mu$. This is not surprising given that Birkhoff's theorem breaks down in GR in the case of scalar fields, \cite{Charmousis:2001nq}. We will see however that even in this restricted case interesting possibilities do arise.

Now going further for arbitrary $p$, if we use \eqref{EOM2} and the fact that $\mathcal{F}$ is closed, then $F_{\sigma_1 ... \sigma_p}$ is only function of $x^\mu$ and verifies,
\beq \label{yannis1}\partial_{[\sigma_1} F_{\sigma_2 ... \sigma_{p+1}]} = 0 \text{ and } \nabla_{\sigma_p}^{(4)} F^{\sigma_1 ... \sigma_p} = 0  \eeq
with respect to the 4 dimensional metric $h_{\rho\sigma}$.
Furthermore, if $p \geq 2$ we can define a new tensor,
\beq J^{\sigma_1 ... \sigma_{p-2}} \dot{=} e^{-2\nu} B^{7/4} F^{\ \ \sigma_1 ... \sigma_{p-2}}_{uv} \eeq which again depends only on the internal coordinates $x^\mu$ and verifies
\beq \label{yannis2}\partial_{[\sigma_1} J_{\sigma_2 ... \sigma_{p-1}]} = 0 \text{ and } \nabla_{\sigma_{p-2}}^{(4)} J^{\sigma_1 ... \sigma_{p-2}} = 0 \ ,\eeq
where $ J_{\sigma_1 ... \sigma_{p-2}} \dot{=} h_{\sigma_1 \rho_1} ... h_{\sigma_{p-2} \rho_{p-2}} J^{\rho_1 ... \rho_{p-2}} $. In the language of differential forms we get the following result: \bs

\textit{
Given the integrability conditions (\ref{uu},\ref{vv}) and metric (\ref{metric}) there exists $\mathcal{F}^{(4)} \in \Lambda^p(\mathcal{H})$ and $\mathcal{J}^{(4)} \in \Lambda^{p-2}(\mathcal{H})$ both closed and co-closed with
\beq \mathcal{F}^{(4)} \dot{=} \frac{1}{p!} F_{\sigma_1 ... \sigma_{p}} \dd{x^{\sigma_1}} \wedge ... \wedge \dd{x^{\sigma_p}}
\text{ and }
\mathcal{J}^{(4)} \dot{=} \frac{1}{(p-2)!} J_{\sigma_1 ... \sigma_{p-2}} \dd{x^{\sigma_1}} \wedge ... \wedge \dd{x^{\sigma_{p-2}}} \ . \eeq
}
$\mathcal{J}^{(4)}$ and $\mathcal{F}^{(4)}$ are what we shall call the electric and  magnetic part of $\mathcal{F}$ respectively. In the electromagnetic case where $p=2$, $\mathcal{J}^{(4)}$ is just a constant function whereas for scalars $\mathcal{J}^{(4)}\equiv 0$. In other words our hypotheses, implementing metric (\ref{metric0}) and integrability conditions (\ref{uu}-\ref{vv}), give a reduction of the bulk $p$-forms which are now living on the 4 space $\mathcal{H}$. We can now move on to the remaining field equations.

The $uv$ left hand side of \eqref{EOM1} reads
\begin{align}
	\label{Euv}
	\mathcal{E}_{uv}
	&= \frac{{B_{,uv} }}{B} - \Lambda e^{2\nu } B^{ - 3/4}  + \frac{\alpha }{2}e^{2\nu } B^{ - 7/4} \hat G^{(4)} \notag \\
	&+ R^{(4)} \left[ {\frac{1}{2}e^{2\nu } B^{ - 5/4}  - \alpha B^{ - 3/2} \left( {\frac{1}{2}\frac{{B_{,u} B_{,v} }}{B} - B_{,uv} } \right)} \right] \notag \\
	&+  \alpha e^{ - 2\nu } B^{ - 5/4} \left[ { - \frac{{15}}{{16}}\left( {\frac{{B_{,u} B_{,v} }}{B}} \right)^2  + \frac{3}{2}\frac{{B_{,u} B_{,v} }}{B}B_{,uv} } \right]
\end{align}
while
\beq T_{uv} = \frac{e^{2\nu}}{2} \left[ \frac{B^\frac{2p-15}{4}}{(p-2)!}\left(J^{(4)}\right)^2 + \frac{B^\frac{-2p-3}{4}}{p!}\left(F^{(4)}\right)^2  \right]  \eeq
where
$\left(J^{(4)}\right)^2 \dot{=} h^{\sigma_1 \rho_1} ... h^{\sigma_{p-2} \rho_{p-2}} J_{\sigma_1 ... \sigma_{p-2}} J_{\rho_1 ... \rho_{p-2}}$ and
$\left(F^{(4)}\right)^2 \dot{=} h^{\sigma_1 \rho_1} ... h^{\sigma_p \rho_p} F_{\sigma_1 ... \sigma_p} F_{\rho_1 ... \rho_p} $ are purely 4-dimensional scalars. 
We also have the $\mu\nu$ equations, for which the left hand side of \eqref{EOM1} can be brought into the form
\begin{align}
	\label{Emunu}
	\mathcal{E}_{\mu \nu}
	&= G_{\mu \nu }^{(4)}  - e^{ - 2\nu } B^{1/4} \left( {\frac{3}{4}B_{,uv}  + 2B\nu _{,uv} } \right)h_{\mu \nu }  + \Lambda B^{1/2} h_{\mu \nu } \notag \\
	&+ \frac{3}{2}\alpha e^{ - 4\nu } \left( {B_{,uu}  - 2\nu _{,u} B_{,u} } \right)\left( {B_{,vv}  - 2\nu _{,v} B_{,v} } \right)h_{\mu \nu } \notag \\
	&-\alpha e^{ - 4\nu } \left[ {\frac{{45}}{{32}}\left( {\frac{{B_{,u} B_{,v} }}{B}} \right)^2  - \frac{{21}}{8}\frac{{B_{,u} B_{,v} }}{B}B_{,uv}  + \frac{3}{2}B_{,uv} ^2   + 3B_{,u} B_{,v} \nu _{,uv} } \right]h_{\mu \nu } \notag \\
	&+ 2 \alpha e^{ - 2\nu } B^{ - 1/4} \left( {\frac{3}{4}\frac{{B_{,u} B_{,v} }}{B} - \frac{1}{2}B_{,uv}  + 4B\nu _{,uv} } \right) G_{\mu \nu }^{(4)} 
\end{align}
while
\beq T_{\mu\nu} = B^\frac{1-p}{2} T_{\mu\nu}\left(\mathcal{F}^{(4)}\right) - B^\frac{p-5}{2} T_{\mu\nu}\left(\mathcal{J}^{(4)}\right) \label{Tmunu} \eeq
where we have defined
\begin{align}
 T_{\mu\nu}\left(\mathcal{F}^{(4)}\right) &= \frac{1}{(p-1)!} h^{\sigma_1 \rho_1} ... h^{\sigma_{p-1} \rho_{p-1}} F_{\mu\sigma_1 ... \sigma_{p-1}} F_{\nu\rho_1 ... \rho_{p-1}} - \frac{1}{2 p!}h_{\mu\nu} \left(F^{(4)}\right)^2 \label{TM} \\
 T_{\mu\nu}\left(\mathcal{J}^{(4)}\right) &= \frac{1}{(p-3)!} h^{\sigma_1 \rho_1} ... h^{\sigma_{p-3} \rho_{p-3}} J_{\mu\sigma_1 ... \sigma_{p-3}} J_{\nu\rho_1 ... \rho_{p-3}} - \frac{1}{2 (p-2)!}h_{\mu\nu} \left(J^{(4)}\right)^2 \ . \label{TE}
\end{align}
Note that  $R^{(4)}$, $G_{\mu\nu}^{(4)}$ and $\hat{G}^{(4)}$ appear in the field equations and will characterise the geometry of $\mathcal{H}$.
Finally, the $(u\rho)$ equation of \eqref{EOM1} gives no information since it is straightforward to check that $T_{u\rho}=0$.

Before attacking the equations of motion let us reiterate a duality symmetry that will permit us to work with 1,2, or 3-forms in $D=6$.
Consider the following map,
\[ \sim
\left\{
	\begin{aligned}
	p																								          &\rightarrow \tilde{p}=6-p \\
	\mathcal{F}^{(4)}\in\Lambda^p\left(\mathcal{H}\right) 		&\rightarrow  \star_{(4)}\tilde{\mathcal{J}}^{(4)}\in\Lambda^{6-\tilde{p}}\left(\mathcal{H}\right) \\
	\mathcal{J}^{(4)}\in\Lambda^{p-2}\left(\mathcal{H}\right) &\rightarrow  \star_{(4)}\tilde{\mathcal{F}}^{(4)}\in\Lambda^{4-\tilde{p}}\left(\mathcal{H}\right) 
	\end{aligned}
\right. \]
This map takes $p$ to $6-p$ forms where $\star_{(4)}$ is the Hodge star operator defined on $\mathcal{H}$. When we apply $\sim$ to the equations of motion, we find the same equations of motion for the tilded quantities. Hence any solution in the presence of $p$-forms is automatically transformed into a solution for $6-p$ forms. To sketch how this comes about, let us begin with the equation of motion for the $p$-form $\mathcal{F}^{(4)}$
\beq \mathrm{d}\mathcal{F}^{(4)} \stackrel{\sim}{\longrightarrow} \mathrm{d}\star_{(4)}\tilde{\mathcal{J}}^{(4)} = 0 \text{ ie } \delta \tilde{\mathcal{J}}^{(4)} = 0 \eeq
\beq \delta \mathcal{F}^{(4)} \stackrel{\sim}{\longrightarrow} \delta\star_{(4)}\tilde{\mathcal{J}}^{(4)} = 0 \text{ ie } \mathrm{d} \tilde{\mathcal{J}}^{(4)} = 0 \ . \eeq
Then, it is similar to iterate this result for $\tilde{\mathcal{F}}^{(4)}$ applying $\sim$ to $\mathcal{J}^{(4)}$. $\mathcal{J}^{(4)}$ and $\mathcal{F}^{(4)}$ are simply interchanged as in the simple EM duality in 4 dimensions. It is straightforward to check that $T_{uv}$ transforms into,
\beq T_{uv} \stackrel{\sim}{\longrightarrow} \frac{e^{2\nu}}{2} \left[ \frac{B^\frac{2\tilde{p}-15}{4}}{(\tilde{p}-2)!}\left(\tilde{J}^{(4)}\right)^2 + \frac{B^\frac{-2\tilde{p}-3}{4}}{\tilde{p}!}\left(\tilde{F}^{(4)}\right)^2  \right]\eeq
since
\begin{align}
\frac{1}{(p-2)!}\left(J^{(4)}\right)^2 \star_{(4)}1 = \mathcal{J}^{(4)}\wedge\star_{(4)}\mathcal{J}^{(4)} \stackrel{\sim}{\longrightarrow}
\star_{(4)}\tilde{\mathcal{F}}^{(4)}\wedge\star_{(4)}\star_{(4)}\tilde{\mathcal{F}}^{(4)}
=&(-1)^{\tilde{p}(4-\tilde{p})}\star_{(4)}\tilde{\mathcal{F}}^{(4)}\wedge\tilde{\mathcal{F}}^{(4)} \notag \\
=&\tilde{\mathcal{F}}^{(4)}\wedge\star_{(4)}\tilde{\mathcal{F}}^{(4)} \notag \\
=& \frac{1}{\tilde{p}!}\left(\tilde{F}^{(4)}\right)^2 \star_{(4)}1
\end{align}
where we have used the fact that $h_{\mu\nu}$ is a riemannian metric and $\star_{(4)}1 \dot{=} \sqrt{h}\mathrm{d}x^1 \wedge ... \wedge \mathrm{d}x^4$ is the volume form  on $\left(\mathcal{H},h\right)$. Finally we study the transformation of $T_{\mu\nu}$ under $\sim$ using \eqref{Tmunu}. In particular,
\beq  F_{\mu \sigma_1 ... \sigma_{p-1}} \stackrel{\sim}{\longrightarrow} \left(\star_{(4)}\tilde{\mathcal{J}}^{(4)}\right)_{\mu\sigma_1 ... \sigma_{5-\tilde{p}}}
= \frac{1}{(\tilde{p}-2)!} \eta_{\mu\sigma_1 ... \sigma_{5-\tilde{p}}}^{\ \ \ \ \ \ \ \ \ \ \ \ \alpha_1 ... \alpha_{\tilde{p}-2}} \tilde{J}_{\alpha_1 ... \alpha_{\tilde{p}-2}} \ .
\eeq
Then, the following identity\footnote{we recall that $\eta_{\sigma_1 ... \sigma_4} = \sqrt{h} \epsilon_{\sigma_1 ... \sigma_4}$}
\beq \eta^{\mu_1 ... \mu_m \sigma_1 ... \sigma_n} \eta_{\nu_1 ... \nu_m \sigma_1 ... \sigma_n} = n!m! \delta_{[\nu_1}^{\mu_1} ... \delta_{\nu_m]}^{\mu_m} \text{ where } m+n=4  \eeq
permits to show that $T_{\mu\nu}$ turns into $\tilde{T}_{\mu\nu}$. In this manner the study of 4 or 5-forms falls into the case of 2 and 1-forms respectively.

\section{Solutions and staticity}
The integrability conditions \eqref{uu} and \eqref{vv} lead to three different classes of solutions, depending on whether the first or the second factor is zero while an additional class emerges for constant $B$ in (\ref{metric}). The corresponding solutions have distinct characteristics and are thus treated separately in what follows. Class I and II are both warped solutions whereas for Class III we have $B=cst$. Class I solutions are only present in Lovelock theory whereas Class II and III are also present in GR theory. Class II solutions in particular give the GR black hole solutions whereas class III contain flat space and unwarped metrics.

\subsection{Class I}

\noindent
Setting the second factor of the equations \eqref{uu} and \eqref{vv} equal to zero leads to the common equation
\beq 1 + \alpha {B^{ - 1/2} R^{(4)}  + \frac{3}{2} \alpha e^{ - 2\nu } B^{ - 5/4} B_{,u} B_{,v} }  = 0\ , \label{classI}\eeq
from which we can solve for the function $\nu(u,v)$ in terms of $B(u,v)$, according to
\beq \nu(u,v) = \frac{1}{2}\ln \left(-\frac{3\alpha }{2}\frac{{B_{,u}B_{,v}}}{B^{5/4}\left(1+\alpha B^{-1/2}R^{(4)}\right)}\right) \ . \label{nu(u,v)}\eeq

Note that this equation constrains the Ricci scalar $R^{(4)}$ of the internal space to be a constant. We are thus required to consider only horizon geometries of a constant scalar curvature as candidate solutions. Substituting the above expression for $\nu(u,v)$ into \eqref{Euv} yields the constraint
\beq B\left(5+12\alpha\Lambda\right) + \alpha^2\left[ \left(R^{(4)}\right)^2 - 6 \hat{G}^{(4)} \right] + 6\alpha\kappa \left[ \frac{B^{\frac{p-4}{2}}}{(p-2)!} \left(J^{(4)}\right)^2 + \frac{B^{\frac{2-p}{2}}}{p!} \left(F^{(4)}\right)^2 \right]= 0 \ .  \label{uvI} \eeq

Then, taking the trace of \eqref{Emunu} and performing the same substitution we end up with the equation
\beq 5+12\alpha\Lambda + 3\alpha\kappa \left[ \frac{(p-4) B^{\frac{p-6}{2}}}{(p-2)!} \left(J^{(4)}\right)^2 + \frac{(2-p) B^{-\frac{p}{2}}}{p!} \left(F^{(4)}\right)^2 \right] =0 \ . \label{trace}\eeq

In the case $p=1$ and $p=3$, it can be shown that since $B$ is not a constant and $h_{\mu\nu}$ is a riemannian metric, $\mathcal{F}^{(4)}=0=\mathcal{J}^{(4)}$ and we get the Born-Infeld limit $5+12\alpha\Lambda=0$. Furthermore, \eqref{uvI} gives the geometrical condition $\hat{G}^{(4)}=\frac{1}{6}\left(R^{(4)}\right)^2$ and we go back to the pure gravitational case of \cite{Bogdanos:2009pc}. So, we conclude that it is impossible to add a free scalar field or a 3-form in the theory for this class of solutions.
	
	

Hence we restrict our attention to the case $p=2$ (and by duality $p=4$) by analysing the $(\mu\nu)$ equations of \eqref{EOM1}. For that, we rewrite \eqref{Emunu} in terms of the trace as
\begin{align}
\mathcal{E}_{\mu\nu} &= \frac{1}{4} B^{1/2} \mathcal{E} h_{\mu\nu}  \notag \\
&+ \left(R^{(4)}_{\mu\nu}-\frac{1}{4}R^{(4)}h_{\mu\nu} \right)
\left[ 1 + 2\alpha e^{-2\nu} B^{-1/4} \left(\frac{3}{4}\frac{{B_{,u} B_{,v} }}{B} - \frac{1}{2}B_{,uv} + 4B\nu_{,uv}\right)\right]
\label{munu}
\end{align}
where $\mathcal{E} = g^{\mu\nu}\mathcal{E}_{\mu\nu}$. Actually, $\mathcal{E}=0$ since the internal part of the stress-energy-momentum tensor is traceless for $p=2$ and $p=4$. Therefore, the $(\mu\nu)$ equation boils down to a nice factorisable form,
\beq
\left(R^{(4)}_{\mu\nu}-\frac{1}{4}R^{(4)}h_{\mu\nu} \right)
\left[ 1 + 2\alpha e^{-2\nu} B^{-1/4} \left(\frac{3}{4}\frac{{B_{,u} B_{,v} }}{B} - \frac{1}{2}B_{,uv} + 4B\nu_{,uv}\right)\right]
=\kappa T_{\mu\nu}\left(\mathcal{F}^{(4)}\right) \ .
\label{munuI}
\eeq

Given the above equation, we have 3 branches of solutions for Class I. For all of them, $\nu(u,v)$ is given by \eqref{nu(u,v)}, $R^{(4)}$ is a constant, we are fine-tuned at the Born-Infeld limit and we have a scalar geometric condition,
\beq
\label{scalar1}
\left(R^{(4)}\right)^2 - 6 \hat{G}^{(4)} + \frac{3\kappa}{\alpha}\left(F^{(4)}\right)^2 = 0 \text{ and } \mathcal{J}^{(4)}=0 \ .  
\eeq
and hence no electric charge is possible for Class I.
The Born-Infeld condition, $5+12\alpha \Lambda=0$ corresponds to a very particular limit in Lovelock theory where the higher order Gauss-Bonnet term is most strongly coupled in comparison to the Einstein-Hilbert term (\ref{action}). This can be explicitely seen by the vacua of Lovelock theory,
\beq 
\mathrm{d}s^2= -V_{vac}(r)\mathrm{d}t^2 + \frac{\mathrm{d}r^2}{V_{vac}(r)} + r^2 \mathrm{d}\Omega_{IV}^2 \ . 
\eeq
where 
\beq
V_{vac}(r)=1+\frac{r^2}{12\alpha}\left(1\pm \sqrt{1+\frac{12\alpha\Lambda}{5}}\right) \ .
\eeq
Note that at the Born-Infeld limit the square root vanishes and the two branches merge into one. The branch with no GR limit (upper + sign) is unstable as long as we are away from the Born Infeld limit where strong coupling occurs. Indeed, at this limit gravitational perturbations around the vacuum are strongly coupled \cite{Charmousis:2008ce} and hence stability is unclear\footnote{Even instanton bounces from one branch to another are very strongly suppresed \cite{Charmousis:2008ce}}.  Furthermore, for $\alpha>0$  and $\Lambda<0$ we have $\alpha\in [0,-\frac{12\Lambda}{5}]$ and hence $\alpha$ attains its maximal value at this limit and we are furthest away from GR. In \cite{fax1} it was shown that degenerate solutions occured for Class I metrics. This is in agreement with the perturbative strong coupling. It is now interesting to see the outcome of these effects in the presence of matter where we can expect to see this degeneracy, partially at least, lifted.
There turn out to be 3 possibilities,
\begin{itemize}
	\item Class Ia: we have that $\mathcal{H}$ is an Einstein space $R^{(4)}_{\mu\nu}=\frac{1}{4}R^{(4)}h_{\mu\nu}$ hence $T_{\mu\nu}=0$. This condition  does not mean that there is no magnetic field present as we will see in the example section. Here, there is no condition on the function $B(u,v)$. This set of solutions is therefore the degenerate class of \cite{fax1}.
	\item Class Ib: Again $T_{\mu\nu}=0$ with the second factor of the left-hand-side of \eqref{munuI} being zero. From \eqref{nu(u,v)}, we get a third order partial differential equation for $B(u,v)$ which reads
\begin{align}
\label{Bfield0}
&\left( {1 + \alpha B^{ - 1/2} R^{\left( 4 \right)} } \right)^2 \left( {B_{,u} ^2 B_{,vv} B_{,uv}  + B_{,v} ^2 B_{,uu} B_{,uv}  - B_{,u} ^2 B_{,v} B_{,uvv}  - B_{,v} ^2 B_{,u} B_{,uuv} } \right) \notag\\
&+ \frac{{B_{,uv} }}{B}B_{,u} ^2 B_{,v} ^2 \left[ {\frac{3}{2} + \frac{5}{2}\alpha B^{ - 1/2} R^{\left( 4 \right)}  + \left( {\alpha B^{ - 1/2} R^{\left( 4 \right)} } \right)^2 } \right] \notag\\
&- \frac{{B_{,u} ^3 B_{,v} ^3 }}{{B^2 }}\left[ {\frac{5}{4} + \frac{{17}}{8}\alpha B^{ - 1/2} R^{\left( 4 \right)}  + \frac{9}{8}\left( {\alpha B^{ - 1/2} R^{\left( 4 \right)} } \right)^2 } \right] = 0 \ .
\end{align}
This equation completely fixes the metric and our solution is no longer degenerate. Note that by continuity we can attain from Class Ib, Class Ia
and therefore impose (\ref{Bfield0}).
	\item Class Ic: There exists a constant $\lambda \neq 0 $ such that
\beq
\lambda\left(R^{(4)}_{\mu\nu}-\frac{1}{4}R^{(4)}h_{\mu\nu}\right) = \kappa T_{\mu\nu}\left(F^{(4)}\right) \text{ for  } p=2 \label{cl1}
\eeq
with $B$ solution of the PDE,
\beq 1 + 2\alpha e^{-2\nu} B^{-1/4} \left(\frac{3}{4}\frac{{B_{,u} B_{,v} }}{B} - \frac{1}{2}B_{,uv} + 4B\nu_{,uv}\right) = \lambda B^{-1/2} \label{Bfield} \ .
\eeq
\end{itemize}
Equation (\ref{Bfield}) is just (\ref{Bfield0}) with the extra matter term. One can find the static solutions by setting $B_{u}=-B_v$ whereupon (\ref{Bfield0}) becomes a second order ODE with respect to $B'$. Making then $r=B^{1/4}$ our radial coordinate we can write the metric as,
\beq
\label{0}
\ud s^2=-U(r) \ud t^2 +24 \frac{\ud r^2}{\alpha R^{(4)}+r^2}+r^2 h_{\mu\nu} \ud  x^\mu \ud x^\nu
\eeq
In particular for $\lambda=0$, the solution reads,
\beq
\label{staticsol}
U(r)=(r^2+\alpha R^{(4)})\left[C_1+C_2\left(\sqrt{\frac{|\alpha R^{(4)}|}{r^2+\alpha R^{(4)}}}- \text{arctanh} \sqrt{\frac{|\alpha R^{(4)}|}{r^2+\alpha R^{(4)}}} \right)\right]^2
\eeq
where $C_1$, $C_2$ are integration constants. 
In the examples section we will construct magnetic solutions to Class I spacetimes by considering distortions of $\mathcal{H}=S^2 \times S^2$. In this case we do not have a local staticity theorem and this is well known \cite{fax1}.


\subsection{Class II}

\subsubsection{Local staticity}
Class-II solutions are obtained by setting (\ref{uu}-\ref{vv}),
\beq  2\nu_{,u}B_{,u}-B_{,uu} = 0 \ (u \leftrightarrow v) \ . \label{int.cond.}\eeq
These integrability conditions are the same as in the case of ordinary GR \cite{bowcock}. We assume that $B$ is not constant. Equation \eqref{int.cond.} implies that
\beq e^{2\nu} = B_{,u} f(v) = B_{,v} g(u) \eeq for arbitrary functions $f$ and $g$, which, in turn, yields $B=B(U+V)$ after the coordinate transformations
\beq (u,v) \longrightarrow U(u)=\int_0^u g(\tilde{u})\mathrm{d}\tilde{u}  \text{ and } V(v)=\int_0^v f(\tilde{v})\mathrm{d}\tilde{v} \ . \eeq
Thus, the metric becomes
\beq \mathrm{d}s^2=- 2 B'(U+V) B^{-3/4} \mathrm{d}U\mathrm{d}V + B^{1/2} h_{\mu\nu} \mathrm{d}x^\mu \mathrm{d}x^\nu \ . \eeq
Perform the following coordinate transformations
\beq 
(U,V) \longrightarrow (\bar{z}=U+V,  \bar{t}=V-U) \longrightarrow (\bar{t},B(\bar{z})) \longrightarrow \left(\bar{t}=t/2,r=B^{1/4}\right) \label{bhmetric}
\eeq
and set 
\beq
\label{clapton}
V(r) \dot{=} -\frac{B'(\bar{z})}{8 r^3},
\eeq
upon which the metric turns into
\beq \mathrm{d}s^2= -V(r)\mathrm{d}t^2 + \frac{\mathrm{d}r^2}{V(r)} + r^2 h_{\mu\nu} \mathrm{d}x^\mu  \mathrm{d}x^\nu \ . \eeq
Class II spacetimes are therefore locally static  since $\partial_t$ is a timelike Killing vector as long as $V>0$. The $(uu)$ and $(vv)$ equations determine the staticity of the metric, as well as the relation between $B$ and $\nu$. This is true here even in the presence of matter as long as $T_{uu}=T_{vv}=0$. We now solve the remaining field equations and characterise the matter fields for each case  in detail.

\subsubsection{General equations for all p}
We can now determine $B$ from the $(uv)$-equation (\ref{Euv}),
\begin{align}
B'' - \Lambda B^{1/4}B' + \frac{\alpha}{2}B^{-3/4}B'\hat{G}^{(4)} + R^{(4)}\left[\frac{1}{2}B^{-1/4}B'+2\alpha\left(B^{1/2}\right)''\right] \notag \\
+ \frac{3\alpha}{4}\left(B^{-5/4}B'^2\right)' -\frac{\kappa}{2}\left[ \frac{B^{\frac{2p-11}{4}}B'}{(p-2)!} \left(J^{(4)}\right)^2 + \frac{B^{\frac{1-2p}{4}}B'}{p!} \left(F^{(4)}\right)^2 \right] = 0
\end{align}
where a prime denotes differentiation with respect to $\bar{z}$. We can integrate the above equation with respect to $\bar{z}$, thus there exists a function $h$, which depends only on the internal coordinates $x^\mu$, such that
\begin{align}
B'- \frac{4\Lambda}{5}B^{5/4} + 2\alpha\hat{G}^{(4)}B^{1/4} + 2R^{(4)}\left[\frac{1}{3}B^{3/4}+\alpha\left(B^{1/2}\right)'\right] + \frac{3\alpha}{4}B^{-5/4}B'^2 \notag\\
+ \frac{2\kappa}{(7-2p)(p-2)!}B^{\frac{2p-7}{4}}\left(J^{(4)}\right)^2 + \frac{2\kappa}{(2p-5)p!}B^{\frac{5-2p}{4}}\left(F^{(4)}\right)^2 = h(x^\mu) \ .
\label{uvII}
\end{align}
	
%

Since $B=B(\bar{z})$, (\ref{uvII}) generically imposes constraints on the horizon geometry $\mathcal{H}$. Furthermore, (\ref{uvII}) will manifestly lead us to a quadratic equation for $B'$ thus determining the potential $V$ from (\ref{clapton}). However, let us first check the $\left(\mu\nu\right)$-equations in order to get the full picture. Using \eqref{munu}, we obtain after some algebra
\begin{align}
\left(R^{(4)}_{\mu\nu} - \frac{1}{4}R^{(4)}h_{\mu\nu}\right)
\left[1+\frac{4\alpha B^{1/4}}{B'}\left(B^{1/2}\frac{U'}{U}\right)'\right]
=& \kappa B^{\frac{1-p}{2}} \left[ T_{\mu\nu}\left(\mathcal{F}^{(4)}\right) -\frac{1}{4} T^{(4)}\left(\mathcal{F}^{(4)}\right) h_{\mu\nu} \right] \notag \\
&- \kappa B^{\frac{p-5}{2}} \left[ T_{\mu\nu}\left(\mathcal{J}^{(4)}\right) -\frac{1}{4} T^{(4)}\left(\mathcal{J}^{(4)}\right) h_{\mu\nu} \right]
\label{munuII}
\end{align}
where $U(z) = -\frac{B'(\bar{z})}{8 B^{3/4}(\bar{z})}$ and $T^{(4)}\left(\mathcal{F}^{(4)}\right) = h^{\rho\sigma}T_{\rho\sigma}\left(\mathcal{F}^{(4)}\right)$ (idem for $\mathcal{J}^{(4)}$). The right hand side matter tensors are given by
\beq T_{\mu\nu}\left(\mathcal{F}^{(4)}\right) -\frac{1}{4} T^{(4)}\left(\mathcal{F}^{(4)}\right) = \frac{1}{(p-1)!}\left[ h^{\sigma_1 \rho_1} ... h^{\sigma_{p-1} \rho_{p-1}} F_{\mu\sigma_1 ... \sigma_{p-1}} F_{\nu\rho_1 ... \rho_{p-1}} - \frac{1}{4}h_{\mu\nu} \left(F^{(4)}\right)^2 \right] \ ,\eeq
\beq T_{\mu\nu}\left(\mathcal{J}^{(4)}\right) -\frac{1}{4} T^{(4)}\left(\mathcal{J}^{(4)}\right) = \frac{1}{(p-3)!}\left[ h^{\sigma_1 \rho_1} ... h^{\sigma_{p-3} \rho_{p-3}} J_{\mu\sigma_1 ... \sigma_{p-3}} J_{\nu\rho_1 ... \rho_{p-3}} - \frac{1}{4}h_{\mu\nu} \left(J^{(4)}\right)^2 \right] \ .\eeq
and are traceless as is the LHS where we have the Einstein spacetime condition.
The trace part of these equations is in fact part of the Bianchi identities and as such, can be obtained from (\ref{uvII}). Any solution of (\ref{uvII}), (\ref{munuII}) and the matter field equations (\ref{yannis1}), (\ref{yannis2}) will be a solution to the ensemble of field equations for all $p$-forms. We now look specifically at each case, $p=1,2,3$ since cases $p=4,5,6$ are deduced by the duality $\sim$.

\subsubsection{The free scalar field}
Here, we have $\mathcal{J}^{(4)} = 0$ and $\mathcal{F}^{(4)}=\partial_\mu \phi \; \mathrm{d} x^{\mu}$, where $\phi$ is a scalar field depending only on the internal coordinates $x^\mu$ as dictated by (\ref{uu}) and (\ref{vv}). Equation (\ref{uvII})
%
implies that $h(x^\mu)=m$ is a constant and we have a quadratic equation for $B'$
	 \beq \frac{3\alpha}{4}B^{-5/4}B'^2 + \left(1+\alpha R^{(4)}B^{-1/2}\right)B' -\frac{4\Lambda}{5}B^{5/4} + \frac{2}{3}\left[R^{(4)}-\kappa\left(F^{(4)}\right)^2\right] B^{3/4} + 2\alpha\hat{G}^{(4)} B^{1/4}- m = 0 \ . \eeq
	 Thus,
	 \beq \label{v1}
	 V(r) = \frac{R^{(4)}}{12} + \frac{r^2}{12\alpha}
	 \left[
	 1\pm\sqrt{1 + \frac{12\alpha\Lambda}{5} + \frac{2\alpha\kappa\left(F^{(4)}\right)^2}{r^2} + \alpha^2\frac{\left(R^{(4)}\right)^2-6\hat{G}^{(4)}}{r^4} + \frac{3\alpha m}{r^5} }
	 \right] \ .
	 \eeq
Since $V=V(r)$, every power of $r$ will have, apart from special cases, a constant coefficient. As such the contribution due to the scalar, $\left(F^{(4)}\right)^2=\partial_\mu\phi \partial^\mu \phi$ is generically a constant. But that is not all. We can write (\ref{munuII}) in a factorisable form,
\beq
\left(R^{(4)}_{\mu\nu} - \frac{1}{4}R^{(4)}h_{\mu\nu}\right)\left[1+\frac{4\alpha B^{1/4}}{B'}\left(B^{1/2}\frac{U'}{U}\right)'\right]
= \kappa\left[F_\mu F_\nu - \frac{1}{4} h_{\mu\nu}\left(F^{(4)}\right)^2\right] 
\eeq
in which the separable nature is clearly manifest. Note that the $\alpha$ dependent terms give now an {\it extra} constraint for the metric potential $V$ which we have already determined in (\ref{v1}). Note furthermore that the $\cal{H}$ dependent terms are  trace free operators of the 4-dimensional metric. Clearly then,  the situation is going to be far more constrained that in GR where $\alpha=0$.
We have two possibilities: 
\begin{itemize}
	\item If $F_\mu F_\nu - \frac{1}{4} h_{\mu\nu}\left(F^{(4)}\right)^2 \neq 0 $ then there exists a separability constant $\lambda \in \mathbb{R}^\star$ such that,
\bea
\lambda\left(R^{(4)}_{\mu\nu} - \frac{1}{4}R^{(4)}h_{\mu\nu}\right) &=& \kappa\left[F_\mu F_\nu - \frac{1}{4} h_{\mu\nu}\left(F^{(4)}\right)^2\right] \ ,\label{ein1}  \\
 1+\frac{4\alpha B^{1/4}}{B'}\left(B^{1/2}\frac{U'}{U}\right)' &=& \lambda \ . \label{pot}
\eea
The constant $\lambda$ is positive from (\ref{ein1}) since the bulk coupling $\kappa/\lambda$ is required positive.
Integrate \eqref{pot} and compare with (\ref{v1}). We then find, 
\beq
\label{bar}
V(r) = \frac{1-\lambda}{12\alpha}r^2 + \rho \text{ with }\rho=\frac{1}{12}\left[R^{(4)} - \frac{\kappa}{\lambda} \left(F^{(4)}\right)^2 \right] 
\eeq
where $\rho$ is constant and the additional constraint on $\mathcal{H}$
\beq
\label{dressing}
\frac{\kappa}{\lambda} \left(F^{(4)}\right)^2 =\sqrt{\left(R^{(4)}\right)^2 - 6\hat{G}^{(4)}}
\eeq
where $\lambda$ is also fixed,
\beq 
5\left(1-\lambda^2\right) + 12\alpha\Lambda = 0 \ . 
\eeq
Using the scalar constraint above and the trace-free part (\ref{ein1}) we can combine the lot to get an effective 4-dimensional Einstein equation with matter,
\beq
\label{ein2}
G^{(4)}_{\mu\nu} +3 \rho h_{\mu\nu}=\frac{\kappa}{\lambda}T_{\mu\nu}^{(4)}(\mathcal{F}^{(4)})
\eeq
where $T_{\mu\nu}^{(4)}(\mathcal{F}^{(4)})$ is the usual 4-dimensional energy-momentum tensor for a scalar field. There are two noteworthy effects here: the scalar field has a net effect of giving a 4-dimensional effective cosmological constant of space $\cal{H}$ given by $3\rho$. The contribution to the 4-dimensional cosmological constant  is enhanced as we approach the Born-Infeld limit where $\lambda\rightarrow 0$. In fact the scalar field permits to move away from the Born-Infeld limit.
Again that is not all. Note that we have also a second order effect in curvature, relating the squared trace of matter with squares of curvature (\ref{dressing}).
Using the geometric identity $2 R^{(4)}_{\mu\nu} - \frac{2}{3}R^{(4)}h_{\mu\nu}=\left(C^{(4)}\right)^2-\hat{G}^{(4)}$, where $\left(C^{(4)}\right)^2$ stands for the square of the Weyl tensor of $\mathcal{H}$, (see for example \cite{ray}) and (\ref{ein1}) we can show that the second order constraint reads,
\beq
\label{teos1}
8\frac{\kappa^2}{\lambda^2} \left[\left(F^{(4)}\right)^2\right]^2 =6 \left(C^{(4)}\right)^2 \ .
\eeq	
Matter, unlike in GR, is related to the conformal Weyl tensor.
We see that on the one hand the internal space $\mathcal{H}$ has to be solution of Einstein's equations in 4-dimensions (\ref{ein2}) but with a conformal dressing given by (\ref{teos1}). This is rather restrictive. For example, take any conformally flat 4-dimensional background with scalar matter. Although an infinity of those can solve (\ref{ein2}) all are excluded in the presence of a free scalar field. If we allow for a non-trivial Weyl tensor we can construct certain solutions as we will see in the example section.
\item If on the other hand $F_\mu F_\nu - \frac{1}{4} h_{\mu\nu}\left(F^{(4)}\right)^2 = 0$, then we will now show that $\mathcal{F}^{(4)}=0$ and there can be no solutions involving a {\it single} scalar field. Indeed, we have $F_\mu F^\nu = \frac{1}{4} \left(F^{(4)}\right)^2 \delta_\mu^\nu$ . Hence,
\beq 
F^{\rho_1} F _{\rho_1} = ... = F^{\rho_4} F _{\rho_4} = \frac{1}{4}\left(F^{(4)}\right)^2 \label{F^2}
\eeq
Moreover, any mixed product is zero for example, $ F_{\rho_1} F^{\rho_2} = 0 $. Hence, we have two possibilities:
\begin{itemize}
\item {$F_{\rho_1}=0$. Then \eqref{F^2} implies that $\left(F^{(4)}\right)^2 = 0$, thus $\mathcal{F}^{(4)}=0$ since $h_{\mu\nu}$ is a riemannian metric}.
\item {Otherwise if $F^{\rho_2}=0$ we arrive at the same conclusion with the same reasoning.}
\end{itemize}
We will show in the example section that taking 4 scalar fields can circumvent this no-go result.
\end{itemize}

\subsubsection{The electromagnetic interaction}
In this case, $\mathcal{J}^{(4)}$ is a constant function corresponding to the electric charge and we have the usual Coulomb electric field strength for 6-dimensional spacetime, $F_{rt} = \mathcal{J}^{(4)}/r^4$. The $(uv)$ equation
implies the quadratic equation,
\beq \frac{3\alpha}{4}B^{-5/4}B'^2 + \left(1+\alpha R^{(4)}B^{-1/2}\right)B' -\frac{4\Lambda}{5}B^{5/4} + \frac{2}{3}R^{(4)} B^{3/4} + 2 a B^{1/4}- m + \frac{2}{3}\kappa\left(J^{(4)}\right)^2 B^{-3/4}= 0  \eeq
and therefore we obtain
	 \bea
&&V(r) = \frac{R^{(4)}}{12} + \nn\\
&&+\frac{r^2}{12\alpha}
	 \left[
	 1\mp\sqrt{1 + \frac{12\alpha\Lambda}{5} + \frac{ W}{r^4} + \frac{3\alpha m}{r^5} - \frac{2\alpha\kappa\left(J^{(4)}\right)^2}{r^8}}
	 \right] \ .\label{pot1}
	 \eea
where 
\beq
\label{miles}
W=\alpha^2 \left(R^{(4)}\right)^2-6\alpha^2\hat{G}^{(4)} + 3\kappa\alpha\left(F^{(4)}\right)^2
\eeq
with $W$ a constant. 
Note how the magnetic and electric field couple differently with the radial coordinate. In fact, the magnetic field changes the falloff behaviour of the solution as $r\rightarrow +\infty$ in particular for $\Lambda=0$. In order to have the same falloff as in asymptotically flat spaces one has to fine-tune the coupling constant $\alpha$. so that $W=0$.
Again, generically all coefficients of powers of $r$, are constant functions, except for particular cases that will be uncovered by examining \eqref{munuII},
\beq
\left(R^{(4)}_{\mu\nu} - \frac{1}{4}R^{(4)}h_{\mu\nu}\right)\left[1+\frac{4\alpha B^{1/4}}{B'}\left(B^{1/2}\frac{U'}{U}\right)'\right]
=  \kappa B^{-1/2} \left[ h^{\rho\sigma}F_{\mu\rho}F_{\nu\sigma} - \frac{1}{4}h_{\mu\nu}\left(F^{(4)}\right)^2 \right] \ .
\eeq
Note that no condition is induced here for the electric part $\mathcal{J}^{(4)}$.
We have all in all three possible cases. Firstly if $h^{\rho\sigma}F_{\mu\rho} F_{\nu\sigma} - \frac{1}{4} h_{\mu\nu}\left(F^{(4)}\right)^2 \neq 0 $ then there exists a constant $\lambda$ such that,
	\begin{align}
	\lambda\left(R^{(4)}_{\mu\nu} - \frac{1}{4}R^{(4)}h_{\mu\nu}\right) &= \kappa\left[h^{\rho\sigma}F_{\mu\rho} F_{\nu\sigma} - \frac{1}{4} h_{\mu\nu}\left(F^{(4)}\right)^2\right] \ , \\
	 1+\frac{4\alpha B^{1/4}}{B'}\left(B^{1/2}\frac{U'}{U}\right)' &= \lambda B^{-1/2} \ . \label{pot2}
	\end{align}
	Integrate \eqref{pot2} to find the following potential,
	\beq V(r) = \frac{r^2}{12\alpha} + p + \frac{q}{2r} - \frac{\lambda}{2\alpha}\ln r \eeq
	where $p$ and $q$ are constants. Comparing with (\ref{pot1}) gives obviously $\lambda=0$.
Therefore we are led to the second possibility whereupon we demand that the traceless part of the magnetic field is precisely zero,  $h^{\rho\sigma}F_{\mu\rho} F_{\nu\sigma} - \frac{1}{4} h_{\mu\nu}\left(F^{(4)}\right)^2 = 0 $ with the magnetic charge $\mathcal{F}^{(4)}\neq 0$. Then
\beq 
\left(R^{(4)}_{\mu\nu} - \frac{1}{4}R^{(4)}h_{\mu\nu}\right)\left[1+\frac{4\alpha B^{1/4}}{B'}\left(B^{1/2}\frac{U'}{U}\right)'\right] = 0 
 \eeq
Annihilating the first factor we have an Einstein space for $\mathcal{H}$,  $R^{(4)}_{\mu\nu} = \frac{1}{4}R^{(4)}h_{\mu\nu} $ with a dyonic black hole potential (\ref{pot1}). This case gives magnetic black hole solutions and has been recently discussed under the condition  that $\mathcal{H}$ is Einstein in \cite{Maeda:2010qz}. The authors there also considered higher order corrections to the magnetic field and we refer the reader for details on this solution. Secondly, $\mathcal{H}$ is not necessarily an Einstein space but	$ V(r) = \frac{r^2}{12\alpha} + p + \frac{q}{2r} $ where $p$ and $q$ are constants. Comparing with  (\ref{pot1}), we have
	\beq q=0 \ , \mathcal{J}^{(4)}=0 ,  p = \frac{1}{12}\left[R^{(4)} \pm \sqrt{\left(R^{(4)}\right)^2 - 6\hat{G}^{(4)} + \frac{3\kappa}{\alpha} \left(F^{(4)}\right)^2 } \right] \text{ and } 5+12\alpha\Lambda=0 \ .\eeq
Therefore $\mathcal{H}$ now verifies only a scalar constraint which is not too surprising given that we are at the Born-Infeld limit.
Note that by combining the above with  a scalar field ($p=1$) we can again move away from the Born-Infeld relation.
Last possibility is to keep only an electric component, $\mathcal{J}^{(4)}$ with zero magnetic charge $\mathcal{F}^{(4)}=0$. Then any Einstein space is a valid 4-dimensional metric with black hole potential,
\beq
V(r) = \frac{R^{(4)}}{12} + \frac{r^2}{12\alpha}
	 \left[
	 1\mp\sqrt{1 + \frac{12\alpha\Lambda}{5} + \frac{ \alpha^2 \left(R^{(4)}\right)^2-6\alpha^2\hat{G}^{(4)} }{r^4} + \frac{3\alpha m}{r^5} - \frac{2\alpha\kappa\left(J^{(4)}\right)^2}{r^8}}
	 \right] \ .\label{pot21}
\eeq
as long as $\hat{G}^{(4)}$ is constant.  For example $S^2\times S^2$ or Bergman space are permissible horizon geometries. 

\subsubsection{3-form matter}
The $(uv)$ equation gives,
%
 \bea
\label{pot31}
&&V(r) = \frac{R^{(4)}}{12} + \nn\\
&&+\frac{r^2}{12\alpha}
\left[
1\mp\sqrt{1 + \frac{12\alpha\Lambda}{5} + \alpha^2\frac{\left(R^{(4)}\right)^2-6\hat{G}^{(4)}}{r^4} + \frac{3\alpha m}{r^5} - \frac{6\alpha\kappa\left[\left(J^{(4)}\right)^2 + \frac{1}{6}\left(F^{(4)}\right)^2 \right]}{r^6} }
\right] 
\eea
Note the self-dual character of the 3-form charges which now couple to the same power of the radial coordinate.
On the other hand \eqref{munuII} gives
\begin{align}
\left(R^{(4)}_{\mu\nu} - \frac{1}{4}R^{(4)}h_{\mu\nu}\right) & \left[1+\frac{4\alpha B^{1/4}}{B'}\left(B^{1/2}\frac{U'}{U}\right)'\right] \notag \\
&= \kappa B^{-1} \left[
\frac{1}{2} h^{\rho\sigma}h^{\alpha\beta}F_{\mu\rho\alpha} F_{\nu\sigma\beta} - \frac{1}{8} h_{\mu\nu}\left(F^{(4)}\right)^2 - J_\mu J_\nu + \frac{1}{4} h_{\mu\nu}\left(J^{(4)}\right)^2 \right] \ .
\end{align}
Note that the 3-form "electric" charge $\mathcal{J}^{(4)}$ is a vector with respect to $\mathcal{H}$ whereas the "magnetic" part is again a 3-form with respect to $\mathcal{H}$.

If the RHS matter sector is non-zero then there exists constant $\lambda \neq 0 $ such that
\begin{align}
\lambda\left(R^{(4)}_{\mu\nu} - \frac{1}{4}R^{(4)}h_{\mu\nu}\right) &= \kappa\left[ \frac{1}{2} h^{\rho\sigma}h^{\alpha\beta}F_{\mu\rho\alpha} F_{\nu\sigma\beta} - \frac{1}{8} h_{\mu\nu}\left(F^{(4)}\right)^2 - J_\mu J_\nu + \frac{1}{4} h_{\mu\nu}\left(J^{(4)}\right)^2 \right] \ , \\
 1+\frac{4\alpha B^{1/4}}{B'}\left(B^{1/2}\frac{U'}{U}\right)' &= \lambda B^{-1} \ . \label{pot3}
\end{align}
Thus, we can again easily integrate \eqref{pot3} and find the following potential
\beq V(r) = \frac{r^2}{12\alpha} + p + \frac{q}{2r} - \frac{\lambda}{4\alpha r^2} \label{pot4}\eeq
where $p$ and $q$ are constants. Then comparing with (\ref{pot31})  we are led to $\lambda=0$!

The only other possibility is to have $\frac{1}{2} h^{\rho\sigma}h^{\alpha\beta}F_{\mu\rho\alpha} F_{\nu\sigma\beta} - \frac{1}{8} h_{\mu\nu}\left(F^{(4)}\right)^2 = J_\mu J_\nu - \frac{1}{4} h_{\mu\nu}\left(J^{(4)}\right)^2$ without switching off the 3-form charges. This turns out to be very restrictive for $\mathcal{F}^{(4)}$ and $\mathcal{J}^{(4)}$ as for the scalar field case. To see this, we introduce $\mathcal{K}^{(4)}=K_\mu \mathrm{d}x^{\mu} \in \Lambda^1\left(\mathcal{H}\right)$ such that $\mathcal{F}^{(4)} = \star_{(4)} \mathcal{K}^{(4)}$. In other words the three form charges correspond to  4-dimensional scalar potentials for $\mathcal{H}$. This takes us back to case $p=1$ with now two scalar fields. We find,
\beq
\frac{1}{2} h^{\rho\sigma}h^{\alpha\beta}F_{\mu\rho\alpha} F_{\nu\sigma\beta} - \frac{1}{8} h_{\mu\nu}\left(F^{(4)}\right)^2 =
- \left[ K_\mu K_\nu - \frac{1}{4} h_{\mu\nu}\left(K^{(4)}\right)^2 \right] \ .
\eeq
Then we have the following relation
\beq J_\mu J_\nu + K_\mu K_\nu = \frac{1}{4} h_{\mu\nu}\left[ \left(J^{(4)}\right)^2 + \left(K^{(4)}\right)^2 \right] \ . \label{J^2,K^2}\eeq
After that, it is not difficult to show that, if $\mu \neq \nu$ then
\beq \left( K^\mu K_\mu + J^\mu J_\mu \right)\left( K^\mu K_\mu - J^\nu J_\nu \right) =0 \text{ (no summation here) } \ . \eeq
We can either have $K^\mu K_\mu + J^\mu J_\mu = 0 $ which implies $\left(J^{(4)}\right)^2 + \left(K^{(4)}\right)^2=0$, hence $\mathcal{F}^{(4)}=0$ and $\mathcal{J}^{(4)}=0$. Or, $\forall \mu \neq \nu$ we have, $K^\mu K_\mu = J^\nu J_\nu$ which implies
\beq K^{\rho_1} K_{\rho_1} = ... = K^{\rho_4} K_{\rho_4} = J^{\rho_1} J_{\rho_1} = ... = J^{\rho_4} J_{\rho_4}\eeq since $dim\left(\mathcal{H}\right) > 2$. \eqref{J^2,K^2} shows that we find the same result as in the $p=1$ case, which permits us to conclude that $\mathcal{F}^{(4)}=0$ and $\mathcal{J}^{(4)}=0$.

We have shown that it is impossible to add a single non-trivial $3$-form in class-II. In a recent paper \cite{Emparan:2010ni}, it was shown that static black holes in $n$-dimensional asymptotically flat spacetime cannot support a non-trivial electric p-form field strengths when $(n+1)/2 \leq  p \leq n-1$ in Einstein theory. Our result confirms their result  for Lovelock theory and shows that-at least for the hypothesis set in our paper- we can go beyond the lower bound of \cite{Emparan:2010ni}. We will however see in the example section that allowing for another 3-form field can actually give a static solution.

\subsection{Class III}
The last class of solutions is given for $B$ constant, set $B \dot{=} \beta^4 \neq 0$. The metric (\ref{metric}) is no longer warped in the internal directions. The field equations \eqref{Euv} and \eqref{Emunu} reduce to
\beq -2\Lambda + \alpha\hat{G}^{(4)}\beta^{-4} + R^{(4)}\beta^{-2} = \frac{\kappa\beta^{2(p-6)}}{(p-2)!}\left(J^{(4)}\right)^2 + \frac{\kappa\beta^{-2p}}{p!}\left(F^{(4)}\right)^2 \label{1}\eeq
\begin{align}
\left(R^{(4)}_{\mu\nu} - \frac{1}{4}R^{(4)}h_{\mu\nu}\right) \left(1+ 8\alpha\beta^3 e^{-2\nu}\nu_{,uv}\right)
=& \kappa \beta^{2(1-p)} \left[ T_{\mu\nu}\left(\mathcal{F}^{(4)}\right) -\frac{1}{4} T^{(4)}\left(\mathcal{F}^{(4)}\right) h_{\mu\nu} \right] \notag \\
&- \kappa \beta^{2(p-5)} \left[ T_{\mu\nu}\left(\mathcal{J}^{(4)}\right) -\frac{1}{4} T^{(4)}\left(\mathcal{J}^{(4)}\right) h_{\mu\nu} \right]
\label{2}\end{align}
where the trace of $\mathcal{E}_{\mu\nu}=\kappa T_{\mu\nu}$ gives
\beq 4\Lambda\beta^2 - R^{(4)} - 8\beta^3 e^{-2\nu} \nu_{,uv} \left(\beta^2+\alpha R^{(4)}\right) = \kappa\beta^{2(1-p)}T^{(4)}\left(F^{(4)}\right) - \kappa\beta^{2(p-5)}T^{(4)}\left(J^{(4)}\right) \ . \label{3}\eeq

The form of (\ref{2}) dictates that in the presence of matter there exists a seperability constant $\lambda \neq 0$ such that
\begin{align}
\lambda\left( R^{(4)}_{\mu\nu} -\frac{1}{4}R^{(4)}h_{\mu\nu} \right) &= \kappa \beta^{2(1-p)} \left[ T_{\mu\nu}\left(\mathcal{F}^{(4)}\right) -\frac{1}{4} T^{(4)}\left(\mathcal{F}^{(4)}\right) h_{\mu\nu} \right] \notag \\
&- \kappa \beta^{2(p-5)} \left[ T_{\mu\nu}\left(\mathcal{J}^{(4)}\right) -\frac{1}{4} T^{(4)}\left(\mathcal{J}^{(4)}\right) h_{\mu\nu} \right] \\
1 + 8\alpha\beta^3 e^{-2\nu} \nu_{,uv} &= \lambda \label{liouville}
\end{align}
From (\ref{liouville}) we see that when $\alpha\neq 0$ and $\lambda \neq 1$, the function $\nu$ obeys a Liouville equation $\nu_{uv}=\frac{\lambda-1}{8\alpha\beta^3}e^{2\nu}$. The latter yields $e^{2\nu}=\frac{8\alpha\beta^3}{\lambda-1}\frac{U'V'}{(U+V)^2}$ for arbitrary functions $U=U(u)$ and $V=V(v)$. Now we can perform the change of coordinates $\left(z=U+V,t=V-U\right)$ and we obtain,
\beq \mathrm{d}s^2 = \frac{4\alpha}{(1-\lambda)z^2} \left( -\mathrm{d}t^2 + \mathrm{d}z^2 \right) + \beta^2 h_{\mu\nu}(x) \mathrm{d}x^\mu \mathrm{d}x^\nu \ . \label{ds1}
\eeq
The curvature of the 2-dimensional spacetime in the $(t-z)$-sections is of constant curvature related to $\lambda$ and $\alpha$.
If on the other hand $\lambda=1$, $\nu_{uv}=0$, then $\nu=f(u)+g(v)$ for some functions $f$ and $g$. Then, we perform the coordinate transformations $U=-\int_0^u e^{2f(x)}\mathrm{d}x$ and $V=\int_0^v e^{2g(x)}\mathrm{d}x$. Finally, the same change of coordinates as before $(U,V)\rightarrow(z,t)$ gives
\beq \mathrm{d}s^2 = \frac{1}{2\beta^3} \left( -\mathrm{d}t^2 + \mathrm{d}z^2 \right) + \beta^2 h_{\mu\nu}(x) \mathrm{d}x^\mu  \mathrm{d}x^\nu \ .
\label{ds2}
\eeq
Therefore Class III solutions are also locally static.
Metric (\ref{ds2}) also coincides with the flat GR solution for $\alpha=0$. In fact taking $\alpha=0$ leads us directly to $\lambda=1$. We recognise in the Wick rotated form of (\ref{ds2}) a Kaluza-Klein metric with 2 extra flat dimensions (\ref{bstrings}). In this sense this class of solutions presents more interest in its Wick rotated form (\ref{bstrings}). 

As for the $h_{\mu\nu}(x)$ metric on the  4-dimensional space $\mathcal{H}$ it has to obey an Einstein equation namely,
\beq
\label{einstein2}
G^{(4)}_{\mu\nu}-\beta^2 \left(\frac{\lambda-1}{4\lambda}-\frac{\Lambda}{\lambda}\right)h_{\mu\nu}=\frac{\kappa}{\lambda} \beta^{2(1-p)} T_{\mu\nu}\left(\mathcal{F}^{(4)}\right)- \frac{\kappa}{\lambda} \beta^{2(p-5)} T_{\mu\nu}\left(\mathcal{J}^{(4)}\right)
\eeq
with the geometrical constraint (\ref{1}). Note that the induced cosmological constant on the 4 dimensional space $\mathcal{H}$ depends on the bulk cosmological constant $\Lambda$ but also on $\lambda$ and $\beta$. Let us examine now the particular cases for each  $p$-form matter source.

\subsubsection{The free scalar field}
We take $\lambda\neq 0$.
Rather than taking a free scalar field here the constant character of $B$ permits us to consider also an arbitrary potential $V(\phi)$ for the scalar field. Indeed take,
\beq
 T_{AB} = \partial_A \phi \partial_B \phi - g_{AB}\left[\frac{1}{2}\partial^C \phi \partial_C \phi + V(\phi) \right] \ .
\eeq
Given the integrability conditions (\ref{uu}-\ref{vv}) the energy-momentum tensor is effectively 4 dimensional since $\phi=\phi(x^\mu)$.
We then have a four dimensional Einstein equation (\ref{einstein2}) with the relevant  $T_{\mu\nu}\left(\mathcal{F}^{(4)}\right)$ for the scalar field. The trace is given by,
\beq 4\left[ \Lambda + \kappa V(\phi) \right]\beta^2-R^{(4)}+\kappa\left(\partial \phi \right)^2  = \frac{\lambda-1}{\alpha}\left(\beta^2+\alpha R^{(4)}\right) \label{constraint1}\eeq and \eqref{1} giving the extra constraint,
\beq
\alpha \hat{G}^{(4)}=2\beta^4 \left[ \Lambda + \kappa V(\phi) \right] -\beta^2  R^{(4)}+  \kappa\beta^2\left(\partial \phi \right)^2 \label{constraint2} \ . \eeq
Given that these spaces have a Kaluza-Klein description in the example section we will examine a cosmological setting. In other words we will consider that $\mathcal{H}$ is a spacetime admitting a lorentzian signature metric.

\subsubsection{The electromagnetic interaction}

This case presents particular interest. For a start (\ref{einstein2}) reduces to,
\beq
\label{ein3}
G^{(4)}_{\mu\nu}-\beta^2 \left(\frac{\lambda-1}{4\lambda}-\frac{\Lambda}{\lambda}+\frac{\kappa\left(J^{(4)}\right)^2}{2\lambda \beta^8}\right)h_{\mu\nu}=\frac{\kappa}{\lambda} \beta^{-2} T_{\mu\nu}\left(\mathcal{F}^{(4)}\right)
\eeq
which reads as a 4-dimensional Einstein equation in the presence of a 4-dimensional  tensor field strength $\mathcal{F}^{(4)}$ with an effective cosmological constant which actually includes the 6 dimensional electric charge. Furthermore, if $h_{\mu\nu}$ is lorentzian the tensor $\mathcal{F}^{(4)}$ can be interpreted as an effective 4-dimensional electromagnetic tensor. 
\eqref{1} implies an additional constraint.
\beq
\label{teos}
\hat{G}^{(4)} - \frac{\kappa}{2\alpha}\left(F^{(4)}\right)^2 =
\frac{4\Lambda\beta^4}{\alpha} - \frac{3R^{(4)}\beta^2}{2\alpha} - \frac{\beta^2}{2\alpha^2}(\lambda-1)\left(\beta^2+\alpha R^{(4)}\right) \ .
\eeq
This last condition implies for example the following: Take $\mathcal{H}$ to be lorentzian signature and consider spherically symmetric solutions of 4-dimensional GR with electromagnetic field $\mathcal{F}^{(4)}$. The only solution is that of Reissner-Nordstrom (with cosmological constant). As such we could construct in GR black string metrics. Here however this solution is disallowed due to the additional scalar constraint (\ref{teos}) which is incompatible with the Reissner-Nordstrom metrics. 
There is however one way out of this. Unlike the case of $p=1$, here, we can switch off the Einstein condition by putting $\lambda=0$ without {\it necessarily setting the 6-dimensional EM tensor to zero}. Then the Einstein equation above reduces to its trace,
\beq
\label{trace2}
4\alpha \Lambda+1=2\kappa\alpha\beta^{-8}\left(J^{(4)}\right)^2
\eeq
where the constant charge $\left(J^{(4)}\right)^2$ permits to avoid fine tuning of $\alpha$ and $\Lambda$. The only dynamical equation is the scalar equation (\ref{teos}). We will study an example of a black string in the example section for $\lambda=0$.

\subsubsection{3-form matter}
We consider $\lambda\neq 0$ as for $p=1$ in order to have a non trivial 3-form energy momentum tensor. For 3-forms we have the relevant Einstein equation (\ref{einstein2}) whose trace is now given by,
\beq 4\Lambda\beta^2-\lambda R^{(4)}-\kappa\beta^{-4}\left[\frac{1}{6}\left(F^{(4)}\right)^2+\left(J^{(4)}\right)^2\right] -\frac{(\lambda-1)\beta^2}{\alpha}=0 \ . 
\eeq
Additionally we have the constraint,
\beq -2\Lambda + \alpha\hat{G}^{(4)}\beta^{-4} + R^{(4)}\beta^{-2} =
\kappa\beta^{-6}\left[\frac{1}{6}\left(F^{(4)}\right)^2+\left(J^{(4)}\right)^2\right] \ .
\label{11}\eeq

\section{Example solutions}
In this section we will construct some example solutions of the 3 classes of spacetimes.
\subsection{Including a magnetic field in class I and II}
Magnetic solutions for $p=2$ can be constructed quite generically by considering $\mathcal{H}=S^2 \times S^2$. The idea is to associate a constant magnetic component supporting each 2-sphere. Let us stick to Class I for definiteness but similar ideas can be applied to other classes (see in particular the magnetic black holes discussed recently in \cite{Maeda:2010qz}).
The 4-dimensional metric of $\mathcal{H}=S^2 \times S^2$ can be written as,
\beq
ds^2=\rho_1^2(d\theta_1^2+\sin^2 \theta_1 d\phi_1^2)+\rho_2^2(d\theta_2^2+\sin^2 \theta_2 d\phi_2^2)
\eeq
where $\rho_1, \rho_2$ are the curvature radii of the 2 spheres. We can remark that,
\beq
R^{(4)}=2\frac{\rho_1^2+\rho_2^2}{\rho_1^2\rho_2^2}, \qquad \hat{G}^{(4)}=\frac{8}{\rho_1^2 \rho_2^2}
\eeq
whereas the magnetic field reads,
\beq
\mathcal{F}^{(4)}=Q_1\sin \theta_1 \mathrm{d}\theta_1 \wedge \mathrm{d}\phi_1+Q_2\sin \theta_2 \mathrm{d}\theta_2 \wedge \mathrm{d}\phi_2 \ .
\eeq
Taking $\rho_1=\rho_2$ and $Q_1=Q_2$ gives that $\mathcal{H}$ is precicely an Einstein space and that $T_{\mu\nu}\left(F^{(4)}\right)=0$ as required by (\ref{munuI}). This is then a Class Ia solution where the $B$ function is an arbitrary function.
Otherwise $\mathcal{H}$ is not an Einstein space but resolves the Class Ic equations (\ref{cl1}) with the geometrical constraint \eqref{scalar1} once we set,
\bea
2\kappa Q_1^2&=\frac{\rho_1^2}{\rho_2^2}\left[(\rho_2^2-\rho_1^2) \lambda+\frac{2\alpha}{3}\frac{12\rho_1^2\rho_2^2-(\rho_1^2+\rho_2^2)^2}{\rho_1^2\rho_2^2}\right]\nonumber\\
2\kappa Q_2^2&=\frac{\rho_2^2}{\rho_1^2}\left[(\rho_1^2-\rho_2^2)\lambda+\frac{2\alpha}{3}\frac{12\rho_1^2\rho_2^2-(\rho_1^2+\rho_2^2)^2}{\rho_1^2\rho_2^2}\right] \ .
\eea
What is interesting to note here is that once $\mathcal{H}$ is not an Einstein space then $B$ is no longer arbitrary, it has to solve (\ref{Bfield}). The fact that $B$ is undetermined is a characteristic of Class Ia solutions in the vacuum \cite{fax1} and hence this class of solutions is degenerate. However, as we see here the addition of matter breaks this degeneracy in a non-perturbative way. In other words even the slightest of difference in the curvature radii $\rho_1, \rho_2$, yields a non trivial change in the spacetime metric. This is typical of strong coupling. This indicates that the degenerate Class Ia solutions are a priori non-physical, the only physical ones being those which are a continuous limit of matter solutions Class Ic and Class Ib. In this way combining solutions of (\ref{Bfield}) and Einstein-spaces $S^2\times S^2$ we can obtain Class I solutions. Matter solutions can be obtained but for illustrative purposes here let us simply take $\mathcal{H}=T^4$ where $R^{(4)}=0$ and $\hat{G}^{(4)}=0$. Taking a static Anzatz for (\ref{cl1}) we obtain the solution,
\beq
\label{statcl1}
\ud s^2=-U(r) \ud t^2 + \frac{\ud r^2}{r^2}+r^2 \delta_{\mu\nu} \ud  x^\mu \ud x^\nu \ . 
\eeq
with
\beq
U(r)=\left(r-\frac{\mu}{r^2}\right)^2
\eeq
Despite appearences, $r_h^3=\mu$ is a curvature singularity if $\mu\neq 0$. The solution  is asymptotically adS. If we take $\alpha R^{(4)}<0$ horizons can be  constructed for $\lambda=0$. 
Note that the static solutions of Class I and Class II do not agree. The vacuum solutions are however the same.

\subsection{Solution in class II in the presence of a scalar field}

Consider class II with a single free scalar field. Moreover let us assume in (\ref{bar}) that the 4-dimensional scalar curvature $\rho=0$ to simplify. Hence the 6-dimensional potential is given by $V(r)=\frac{1-\lambda}{12\alpha}r^2$ whereas the Einstein equation for the internal space $\mathcal{H}$ becomes $R^{(4)}_{\mu\nu}=\frac{\kappa}{\lambda}\partial_\mu\phi\partial_\nu\phi$ where $\phi$ is  harmonic on $\mathcal{H}$. We want to examine static spherically metrics whose Lorentzian version are given in  \cite{Agnese:1985xj},
\beq \mathrm{d}s^2 = \left(1-\frac{2\eta}{R}\right)^{\cos\chi}\mathrm{d}\tau^2 + \frac{\mathrm{d}R^2}{\left(1-\frac{2\eta}{R}\right)^{\cos\chi}} + \left(1-\frac{2\eta}{R}\right)^{1-\cos\chi}R^2\left(\mathrm{d}\theta^2+ \sin^2\theta\mathrm{d}\phi^2\right)\eeq
with the scalar field
\beq \phi = \sqrt{\frac{\lambda}{2\kappa}}\sin\chi\ln\left(1-\frac{2\eta}{R}\right) \ . \eeq
In particular, we have the Schwarzschild solution for $\chi=0$ (whereas  $\chi=\pi/3$ is conformally related to the BBMB solution \cite{Bekenstein:1975ts, Bekenstein:1974sf,BBM}). Here we have the additional scalar constraint $\left(C^{(4)}\right)^2 = \frac{4\kappa^2}{3\lambda^2}\left(\partial_\mu\phi\partial^\mu\phi\right)^2$ which is only satisfied for $\chi=\pi/2$:
\beq \mathrm{d}s^2 = \mathrm{d}\tau^2 + \mathrm{d}R^2 + \left(1-\frac{2\eta}{R}\right)R^2\left(\mathrm{d}\theta^2+ \sin^2\theta\mathrm{d}\phi^2\right)\eeq with
\beq \phi = \sqrt{\frac{\lambda}{2\kappa}}\ln\left(1-\frac{2\eta}{R}\right) \ . \eeq
This solution is singular when $R$ tends to $2\eta^+$ since $R^{(4)}=\frac{2\eta^2}{(2\eta-R)^2 R^2}$. In fact point $\chi=\pi/2$ is when we are furthest away from the GR black hole.

\subsection{Black hole solution with two 3-forms on the 4-torus}

As we saw, for Class II metrics, we cannot have a  static black hole solution sourced by a 3-form. However, given that $\mathcal{H}$ is four dimensional take rather two 3-forms in the theory:
\beq 
S^{(6)} = \frac{M^{(6)^4}}{2} \int_\mathcal{M} \mathrm{d}^6 x \sqrt{-g^{(6)}} \left[ R - 2\Lambda + \alpha \hat{G} - \frac{\kappa_1}{6} F_{(1)ABC}F_{(1)}^{ABC} - \frac{\kappa_2}{6} F_{(2)ABC}F_{(2)}^{ABC} \right] \ .
\eeq
Applying the same method as before we find that the 3-forms must imperatively satisfy the following matter condition,
\beq
\label{sum1}
\sum_{i=1}^2 \kappa_i \left[\frac{1}{2} h^{\rho\sigma}h^{\alpha\beta}F_{(i)\mu\rho\alpha} F_{(i)\nu\sigma\beta} - \frac{1}{8} h_{\mu\nu}\left(F_{(i)}^{(4)}\right)^2 - J_{(i)\mu} J_{(i)\nu} + \frac{1}{4} h_{\mu\nu}\left(J_{(i)}^{(4)}\right)^2 \right]= 0 \ .
\eeq 

Effectively each three-form boils down to 2 free scalar field potentials. This can be seen by introducing,
for each 
$i=1,2$, $\mathcal{K}_{(i)}^{(4)}=K_{(i)\mu}\mathrm{d}x^\mu \in \Lambda^1(\mathcal{H})$ such that $\mathcal{F}_{(i)}^{(4)} = \star_{(4)} \mathcal{K}_{(i)}$. 
Hence \eqref{sum1} becomes
\beq \sum_{i=1}^2 \kappa_i \left[J_{(i)\mu} J_{(i)\nu} + K_{(i)\mu} K_{(i)\nu} \right] = \frac{1}{4} h_{\mu\nu} \sum_{i=1}^2 \kappa_i\left[ \left(J_{(i)}^{(4)}\right)^2 + \left(K_{(i)}^{(4)}\right)^2 \right] \ . \label{sum2} \eeq
It is now clear that the simplest of Einstein spaces, $\mathcal{H}=T^4 = S^1 \times S^1 \times S^1 \times S^1 $ is a valid horizon geometry satisfying (\ref{sum2}). Indeed if $(x,y,z,w)$ are the coordinates on $T^4$, we can choose
\beq \mathcal{J}_{1}^{(4)} = \frac{Q}{\sqrt{\kappa_1}}\mathrm{d}x \ ; \mathcal{K}_{1}^{(4)} = \frac{Q}{\sqrt{\kappa_1}}\mathrm{d}y \ ; \mathcal{J}_{2}^{(4)} = \frac{Q}{\sqrt{\kappa_2}}\mathrm{d}z \text{ and } \mathcal{K}_{2}^{(4)} = \frac{Q}{\sqrt{\kappa_2}}\mathrm{d}w \eeq or equivalently
\beq \mathcal{J}_{1}^{(4)} = \frac{Q}{\sqrt{\kappa_1}}\mathrm{d}x \ ; \mathcal{F}_{1}^{(4)} = \frac{Q}{\sqrt{\kappa_1}}\mathrm{d}x\wedge\mathrm{d}z\wedge\mathrm{d}w \ ; \mathcal{J}_{2}^{(4)} = \frac{Q}{\sqrt{\kappa_2}}\mathrm{d}z \text{ and } \mathcal{F}_{2}^{(4)} = \frac{Q}{\sqrt{\kappa_2}}\mathrm{d}x\wedge\mathrm{d}y\wedge\mathrm{d}z \eeq
in order to verify \eqref{sum2} with the arbitrary charge $Q$ being  a constant. In fact each of the four 3-form components is switched on in a different direction of $\mathcal{H}$. As such the whole configuration on $\mathcal{H}$ remains homogeneous.
At the end we have a black hole solution which supports two non-trivial 3-forms with the metric:
\beq 
\mathrm{d}s^2 = - V(r)\mathrm{d}t^2 + \frac{\mathrm{d}r^2}{V(r)} + r^2\left(\mathrm{d}x^2+\mathrm{d}y^2+\mathrm{d}z^2+\mathrm{d}w^2\right)\eeq
with \beq  V(r)=\frac{r^2}{12\alpha}\left[1\pm\sqrt{1+\frac{12\alpha\Lambda}{5}+\frac{3\alpha m}{r^5}-\frac{24\alpha Q^2}{r^6}}\right]
\eeq
This black hole solution is asymptotically locally adS and has a GR limit.
In the Einstein limit, $\alpha\rightarrow 0$, the potential reads  $V(r)=-\frac{\Lambda}{10}r^2-\frac{m}{8}\frac{1}{r^3}+\frac{Q^2}{r^4}$ where we chose the minus sign in the solution since the plus sign is unstable \cite{Charmousis:2008ce}. This is to our knowledge the first static black hole solution involving 3-forms in GR or Lovelock theory. The solution has similar structure to the 6 dimensional version of the Reissner-Nordstrom
planar black hole solution. It presents particular interest and we will come back to it in a future study.

\subsection{Black hole solution with four free scalar fields on the 4-torus}
In complete analogy with the 3-form example given above, we can construct a static solution with four free scalar fields of the action:
\beq S^{(6)} = \frac{M^{(6)^4}}{2} \int_\mathcal{M} \mathrm{d}^6 x \sqrt{-g^{(6)}} \left[ R - 2\Lambda + \alpha \hat{G} - \sum_{i=1}^4\kappa_i \partial_A\phi_{(i)}\partial^A\phi_{(i)} \right]  \ . \eeq
Taking $\mathcal{H}$ to be Euclidean space the 6-dimensional solution reads,
\beq \mathrm{d}s^2 = - V(r)\mathrm{d}t^2 + \frac{\mathrm{d}r^2}{V(r)} + r^2\left(\mathrm{d}x^2+\mathrm{d}y^2+\mathrm{d}z^2+\mathrm{d}w^2\right)\eeq
with \beq  V(r)=\frac{r^2}{12\alpha}\left[1\pm\sqrt{1+\frac{12\alpha\Lambda}{5} + \frac{8\alpha\lambda^2}{r^2} + \frac{3\alpha m}{r^5}}\right]  \eeq
and \beq \phi_{(1)}=\frac{\lambda x}{\sqrt{\kappa_1}} \ ; \phi_{(2)}=\frac{\lambda y}{\sqrt{\kappa_2}} \ ; \phi_{(3)}=\frac{\lambda z}{\sqrt{\kappa_3}} \text{ and } \phi_{(4)}=\frac{\lambda w}{\sqrt{\kappa_4}}  \eeq
where the scalar charge $\lambda$ is a constant. It is interesting to take the Einstein limit whereupon we find, 
\beq
V(r)=-\frac{\Lambda}{10}r^2-\frac{\lambda^2}{3}-\frac{m}{8}\frac{1}{r^3} \ .
\label{hair1}
\eeq
Again we will come back to this solution in a further publication but note that compactification of $\mathcal{H}$ may lead to distributional singularities for the scalars. Also although the horizon space is flat the geometry is that of a hyperbolic black hole. Lastly, we can easily upgrade the solution involving both scalars and 3-forms with the potential,
\beq
V(r)=\frac{r^2}{12\alpha}\left[1\pm\sqrt{1+\frac{12\alpha\Lambda}{5} + \frac{8\alpha\lambda^2}{r^2} + \frac{3\alpha m}{r^5}-\frac{24 \alpha Q^2}{r^6}}\right] \ .
\eeq

\subsection{General solution in class II with all p-forms}
As a final example we can consider the generic case involving  a scalar field, an electromagnetic interaction and a $3$-form given by the action
\beq S^{(6)} = \frac{M^{(6)^4}}{2} \int_\mathcal{M} \mathrm{d}^6 x \sqrt{-g^{(6)}} \left[ R - 2\Lambda + \alpha \hat{G} - \kappa_1 F_{A}F^{A} - \frac{\kappa_2}{2} F_{AB}F^{AB} - \frac{\kappa_3}{6} F_{ABC}F^{ABC} \right] \ .
\eeq
The point we would like to make here is that combination of matter forms can lead to a more generic potential $V$. This comes about as follows:
as we saw for Class II solutions  we can initially express the potential $V$ from the $(uv)$ equation (\ref{uvII}). 
Indeed defining $p,q$ and $t$ such that $\alpha \hat{G}^{(4)}-\frac{\kappa_2}{2}F^2 = p$, $\kappa_1\left(\partial\phi\right)^2= q$ and $\kappa_3\left(J^2+\frac{1}{6}H^2 \right)= t$ and we have,
\beq
V(r) = \frac{R^{(4)}}{12} + \frac{r^2}{12\alpha}
\left[
1\mp\sqrt{1 + \frac{12\alpha\Lambda}{5} + \frac{2\alpha q}{r^2} + \frac{\alpha^2\left(R^{(4)}\right)^2-6\alpha p}{r^4} + \frac{3\alpha m}{r^5} - \frac{6\alpha t}{r^6} - \frac{2\alpha\kappa_2 Q^2}{r^8}}
 \right]
\label{complex1}
\eeq
where $m$ is a constant. This is just a necessary condition for $V$. Using on the other hand the $(\mu\nu)$ equations (\ref{munuII}) we have a condition for the traceless part of the metric tensor on $\mathcal{H}$,
\beq
S^{(4)}_{\mu\nu}\left[1+\frac{4\alpha B^{1/4}}{B'}\left(B^{1/2}\frac{U'}{U}\right)'\right]= \kappa_1\bar{T}_{\phi\mu\nu} + \kappa_2 B^{-1/2} \bar{T}_{F\mu\nu} + \kappa_3 B^{-1} \left(\bar{T}_{J\mu\nu}+\bar{T}_{H\mu\nu}\right)
\eeq
where in turn, $T_{\phi\mu\nu}$,$T_{F\mu\nu}$,$T_{H\mu\nu}$ are given by \eqref{TM} with $p=1,2,3$ respectively, $T_{J\mu\nu}$ by \eqref{TE} with $p=3$ and $S^{(4)}_{\mu\nu}=R^{(4)}_{\mu\nu}-\frac{1}{4}R^{(4)}h_{\mu\nu}$. Here we have denoted by $\bar{T}_{\phi\mu\nu}$ the traceless part of $T_{\phi\mu\nu}$ and so forth. Then, after two successive integrations with respect to $r$, there exists two traceless symmetric tensors $X_{\mu\nu}$ and $Y_{\mu\nu}$ such that
\beq 
\label{complex2}
S^{(4)}_{\mu\nu}V(r) = \frac{S^{(4)}_{\mu\nu}-\kappa_1\bar{T}_{\phi\mu\nu}}{12\alpha}r^2 - \frac{\kappa_2}{2\alpha}\bar{T}_{F\mu\nu}\ln r + \frac{X_{\mu\nu}}{8\alpha}\frac{1}{r} - \frac{\kappa_3}{4\alpha}\left(\bar{T}_{J\mu\nu}+\bar{T}_{H\mu\nu}\right)\frac{1}{r^2}-\frac{1}{32\alpha}Y_{\mu\nu} \ . 
\eeq
Now we need to compare (\ref{complex1}) and (\ref{complex2}).
Assuming $\alpha<0$, $ce>0$ we can summarize the conditions for a full solution such that $S_{\mu\nu}^{(4)}\neq 0$ as follows: 
\bs
\BC
\fbox{\begin{minipage}{14cm}
$$ V(r)=\frac{1-c}{12\alpha}r^2 + \frac{a}{4\alpha} + \frac{e}{4\alpha r^2} \text{ with } c=\pm\sqrt{1+\frac{12\alpha\Lambda}{5}} \ ;  e=\pm\sqrt{\frac{-2\alpha\kappa_2}{9}}|Q| $$
$$  h^{\rho\sigma}F_{\mu\rho}F_{\nu\sigma} = \frac{1}{4}h_{\mu\nu}\left(F^{(4)}\right)^2 $$
$$G^{(4)}_{\mu\nu} + \frac{3a}{4\alpha} h_{\mu\nu} = \frac{\kappa_1}{c}T_{\phi\mu\nu} = -\frac{\kappa_3}{e}\left(T_{H\mu\nu}+T_{J\mu\nu}\right) $$
$$\left(C^{(4)}\right)^2 = \frac{4\kappa_1^2}{3c^2}\left[(\partial\phi)^2\right]^2 + \frac{\kappa_2}{2\alpha}\left(F^{(4)}\right)^2 + \frac{ec}{\alpha^2} $$
\end{minipage} } 
\EC
Moreover each $p$-form has to solve its proper equations of motion. The value of $a$ is actually determined by the trace of the Einstein equation in the box,
\beq
a=\frac{\alpha}{3}\left(R^{(4)} - \frac{\kappa_1}{c}\left(\partial\phi\right)^2\right) \ .
\eeq

\subsection{Black string in Class III}

Consider an electric charge $J^{(4)}$ emanating from an EM tensor, of $p=2$ in Class III with zero magnetic field $\mathcal{F}^{(4)}=0$. Take separability constant $\lambda=0$ in (\ref{liouville}) and $\beta=1$. If we double Wick rotate the 6-dimensional metric (\ref{bstrings}) we have a Kaluza-Klein spacetime with two extra curved directions (\ref{ds1}). Therefore the internal space $\mathcal{H}$ is a Lorentzian 4-dimensional spacetime. Consider a simple Anzatz of a spherically symmetric spacetime so that the metric on $\mathcal{H}$ can be written,
\beq
\ud s^2_4=-f(r) \ud t^2+\frac{\ud r^2}{f(r)}+r^2\left( \frac{\mathrm{d}\chi^2}{1-\kappa\chi^2} + \chi^2\mathrm{d}\theta^2 \right) \ .
\eeq
From (\ref{trace2}) we obtain a simple relation inbetween $\alpha$, $\Lambda$ and $J^{(4)}$. All we have to solve is (\ref{teos}) which gives,
\beq
f(r)=\kappa+\frac{r^2}{4\alpha}\left(1\pm \sqrt{\frac{4}{3}(1+2\alpha\Lambda)+\frac{\alpha^{3/2} \mu}{r^3}-\frac{\alpha^2 q}{r^4} + \frac{16\alpha^2\kappa^2}{r^4} }\right) \ .
\eeq
This solution reduces to the one discussed recently by \cite{mad2}, \cite{mad} when we set $4\alpha \Lambda=-1$, in other words here, with the inclusion of the EM tensor charge, we avoid-at least one-fine tuning. However, do note that once $\lambda\neq 0$ we have to solve the reduced Einstein equation (\ref{ein3}) which is incompatible with the scalar constraint. Furthermore, if we input in $\mathcal{H}$ the most general spherically symmetric Anzatz we will have a solution with one undetermined metric function for $\lambda=0$. The problem here originates in the fact that a scalar metric equation does not possess gauge degrees of freedom. Nevertheless, on the positive side the solution does have an intriguing property. Taking the small $\alpha$ limit and setting $\kappa J^2=\frac{1}{4\alpha}$ we obtain a solution which approximates the 4-dimensional Reissner-Nordstrom solution in ordinary GR. 

\subsection{EGB cosmology in class III}

Consider now a scalar field ($p=1$) with some potential term $V(\phi)$ in a Kaluza-Klein double Wick rotated metric (\ref{bstrings}) for $\beta=1$ and Class III. 
Here we want to study a cosmological 4 dimensional spacetime $\mathcal{H}$ with some perfect fluid source originating from the scalar field. Hence we begin with,
\beq S^{(6)} = \frac{M^{(6)^4}}{2} \int_\mathcal{M} \mathrm{d}^6 x \sqrt{-g^{(6)}} \left[ R - 2\Lambda + \alpha \hat{G} - \kappa\left(\partial_A\phi \partial^A \phi + 2V(\phi) \right) \right] \eeq
where field equations are
\beq G_{AB} + \Lambda g_{AB} - \alpha H_{AB} = \kappa T_{AB} \eeq
with the stress-energy-momentum tensor
\beq T_{AB} = \partial_A \phi \partial_B \phi - g_{AB}\left[\frac{1}{2}\partial^C \phi \partial_C \phi + V(\phi) \right] \eeq
and we have the Klein-Gordon equation $\Box\phi=V'(\phi)$. We will consider the case $\lambda=1$ where the transverse space is flat{\footnote{For this set-up we can consider codimension 2 junction conditions for a perfect fluid source although here for clarity we stick to regular metrics.}} and we have a Kaluza-Klein cosmology with 2 extra dimensions. Thus the metric is given by (\ref{ds2}),
and we have a 4-dimensional  Einstein equation
\beq G_{\mu\nu} + \Lambda g_{\mu\nu} = \kappa\left[\partial_\mu \phi \partial_\nu \phi - h_{\mu\nu}\left[\frac{1}{2}\partial^\rho \phi \partial_\rho \phi + V(\phi) \right]\right]\eeq
with the additional constraint $\alpha\hat{G}^{(4)}+2\left[\Lambda+\kappa V(\phi)\right]=0$ originating from (\ref{constraint2}). Clearly we can input the cosmological constant in the potential but to keep up with previous notation we leave as is. 
We can rewrite the field equation under a form where the dependence with $\Lambda$ and the potential is dropped:
\beq G_{\mu\nu} - \frac{\alpha}{2}\hat{G}^{(4)} h_{\mu\nu}= \kappa\left[\partial_\mu \phi \partial_\nu \phi - \frac{1}{2} h_{\mu\nu}\partial^\rho \phi \partial_\rho \phi \right] \ . \label{cosmo} \eeq
Here we see that the potential term is replaced by the 4-dimensional Gauss-Bonnet scalar.
Now an LFRW-type ansatz gives for $\mathcal{H}$
\beq 
\ud s^2=-\mathrm{d}t^2 + a^2(t)\left(\mathrm{d}x^2+\mathrm{d}y^2+\mathrm{d}z^2\right) 
\eeq 
where $a(t)$ is the scale factor. Hence \eqref{cosmo} gives two equations
\bea
\label{rory1}
3 H^2 &=& \frac{\kappa}{2}\dot{\phi}^2 - 12\alpha H^2 \frac{\ddot{a}}{a} \\
-H^2-2\frac{\ddot{a}}{a} &=& \frac{\kappa}{2}\dot{\phi}^2 + 12\alpha H^2 \frac{\ddot{a}}{a}
\label{rory2}
\eea
where we have introduced the Hubble parameter $H=\dot{a}/a$ 
and we have assumed that the scalar field depends only on $t$ in accord with the 4-dimensional cosmological symmetries. 
Defining in the usual way $\rho=\frac{1}{2}\dot{\phi}^2 - 12\frac{\alpha}{\kappa} H^2 \frac{\ddot{a}}{a}$ and $P=\frac{1}{2}\dot{\phi}^2 + 12\frac{\alpha}{\kappa} H^2 \frac{\ddot{a}}{a}$, we recognize two Friedmann equations,
\bea
H^2 &=& \frac{\kappa}{3}\rho \label{H3} \ ,\\
\frac{\ddot{a}}{a} &=& -\frac{\kappa}{6}\left(\rho+3P\right) \ .
\eea
The matter equation of state between $\rho$ and $P$ is fixed by,
\beq \left(1+4\alpha\kappa\rho\right)P = \left(1-\frac{4}{3}\alpha\kappa\rho\right)\rho  \ . \eeq
It is interesting to treat the scalar derivative as our source while  combining (\ref{rory1}) and (\ref{rory2}). In this way we obtain the modified Friedmann equation which has two branches and is given by,
\beq
H^2=\frac{1}{8\alpha}\left(2\alpha\kappa \dot\phi^2 -1\pm \sqrt{1-\frac{4\alpha}{3}\kappa\dot\phi^2+4\alpha^2\kappa^2 \dot\phi^4}\right) \ .
\eeq

\section{Conclusions}
In this study we considered gravitational solutions of Lovelock theory in 6 dimensions involving $p$-form  matter sources (\ref{action}). Our hypotheses included a quite generic metric Anzatz (\ref{metric}) involving an arbitrary four dimensional space $\mathcal{H}$, internal metric and time and space dependent warp factors (or again (\ref{bstrings})). Our principal assumption involved the matter sector where we imposed the validity of two conditions (\ref{uu}), (\ref{vv}) that had given in a previous study, \cite{Bogdanos:2009pc}, the general solution in the absence of sources. For the hypotheses above, we again demonstrated the integrability of the gravitational problem in the sense that all solutions can be completely classified in 3 classes  with explicit and very restrictive conditions on the internal metric of $\mathcal{H}$. In fact, the situation is far more stringent than in the analogue GR analysis where any Einstein metric on $\mathcal{H}$ can give rise to 6-dimensional solutions. 

We saw that Class I solutions are degenerate and involve the strongly coupled Born-Infeld limit of the theory. Degenerate, for in some cases metric functions are indetermined due to the absence of field equations in this particular limit \cite{fax1}, \cite{Charmousis:2008ce}. We found that reduction of symmetry however, allows for even this class to have fixed non-degenerate solutions. One is tempted to actually take only these into account and discard all other degenerate solutions in this class. Class II solutions are shown to be verifying a local staticity theorem, even in the presence of sources, and involve black hole metrics under certain conditions. The solutions here include known static solutions of this theory such as those of Boulware and Deser \cite{Boulware}, Cai \cite{Cai}, \cite{cai2} and those found more recently in \cite{Dotti}, \cite{Bogdanos:2009pc}, \cite{Maeda:2010qz}. More importantly we have in fact shown that if we restrict solutions to only those which are asymptotically flat, then, only the Boulware-Deser \cite{Boulware} black hole is an allowed solution to which one can include also electric charge. An exception to this rule is permitted by including a magnetic field and fine tuning the Gauss-Bonnet coupling constant $\alpha$ so that $W=0$ in (\ref{miles}). Apart from this case involving fine-tuning the situation is in very close accord with the 4-dimensional GR version of Birkhoff's theorem. Note, that this is not true in higher dimensional GR where Einstein spaces are permissible as horizon metrics. This fact seems to agree with our intuition that the classical higher dimensional version of GR is in fact Lovelock theory and the degeneracy appearing in the horizon metrics is due to the fact that we do not take into account the full 2nd order field equations. Class III solutions also verify a local staticity theorem and are codimension-2 (unwarped) Kaluza-Klein compactifications of $\mathcal{H}$. The solutions we have encountered involve as limits some known solutions of Lovelock theory such as those found previously by \cite{mad,mad2}. We see that such solutions are again degenerate although there is less fine tuning involved once we allow for the presence of matter fields. 

We then went on to find specific example solutions within these classes in order to get an overall picture of some of their properties. We saw that in Class I novel geometries can be constructed which are not degenerate. For Class II the very stringent requirements on $\mathcal{H}$ led us to find scalar and 3-form static solutions which can describe, black hole spacetimes. These solutions are new to 6-dimensional GR and are to our knowledge the first example of static black hole solutions involving 3-form fields. They are asymptotically locally adS and their black hole horizon is flat. We also found in Class III a static 6-dimensional black string metric and 6-dimensional Kaluza-Klein cosmology with a time dependent scalar field. We found that what is percieved as a certain $p$-form matter in higher dimensions may have a completely different interpretation in four dimensions. For example, we saw that the Coulomb electric charge played the role of an effective cosmological constant whereas the scalar field could result in modifying the horizon curvature term in a black hole potential. Certain solutions we have touched upon here certainly ask for further investigation which we hope to report on in the near future.

\section*{Acknowledgements}

It is a great pleasure to thank Lorenzo Sorbo, the CPT Durhamites for encouraging remarks and corrections, Marco Caldarelli, Dr. Blaise Gout\'eraux, Tony H Padilla and Robin Zegers. This work was partially supported by the ANR grant STR-COSMO, ANR-09-BLAN-0157. TK thanks the Laboratory of Theoretical Physics of Orsay for hospitality where part of this work was carried out. CC thanks R-Y Cai, Jon Shock and B Wang for they hospitality at the KITPC where part of this research was carried out.

\bibliography{MatterEGB}

\providecommand{\href}[2]{#2}\begingroup\raggedright\begin{thebibliography}{10}

\bibitem{Will}
C.~M. Will, {\it {The confrontation between general relativity and
  experiment}},  {\em Living Rev. Rel.} {\bf 9} (2005) 3,
  [\href{http://xxx.lanl.gov/abs/gr-qc/0510072}{{\tt gr-qc/0510072}}].

\bibitem{Brans}
C.~Brans and R.~H. Dicke, {\it {Mach's principle and a relativistic theory of
  gravitation}},  {\em Phys. Rev.} {\bf 124} (1961) 925--935.

\bibitem{Damour}
T.~Damour and G.~Esposito-Farese, {\it {Tensor multiscalar theories of
  gravitation}},  {\em Class. Quant. Grav.} {\bf 9} (1992) 2093--2176.

\bibitem{Gilles}
G.~Esposito-Farese, {\it {Tests of scalar-tensor gravity}},  {\em AIP Conf.
  Proc.} {\bf 736} (2004) 35--52,
  [\href{http://xxx.lanl.gov/abs/gr-qc/0409081}{{\tt gr-qc/0409081}}].

\bibitem{lovelock}
D.~Lovelock, {\it {The Einstein tensor and its generalizations}},  {\em J.
  Math. Phys.} {\bf 12} (1971) 498--501.

\bibitem{Zumino}
B.~Zumino, {\it {Gravity Theories in More Than Four-Dimensions}},  {\em Phys.
  Rept.} {\bf 137} (1986) 109.

\bibitem{Deruelle}
N.~Deruelle and J.~Madore, {\it {On the quasi-linearity of the Einstein-
  'Gauss-Bonnet' gravity field equations}},
  \href{http://xxx.lanl.gov/abs/gr-qc/0305004}{{\tt gr-qc/0305004}}.

\bibitem{charm}
C.~Charmousis, {\it {Higher order gravity theories and their black hole
  solutions}},  {\em Lect. Notes Phys.} {\bf 769} (2009) 299--346,
  [\href{http://xxx.lanl.gov/abs/0805.0568}{{\tt arXiv:0805.0568}}].

\bibitem{Garraffo}
C.~Garraffo and G.~Giribet, {\it {The Lovelock Black Holes}},  {\em Mod. Phys.
  Lett.} {\bf A23} (2008) 1801--1818,
  [\href{http://xxx.lanl.gov/abs/0805.3575}{{\tt arXiv:0805.3575}}].

\bibitem{Zwiebach}
B.~Zwiebach, {\it {Curvature Squared Terms and String Theories}},  {\em Phys.
  Lett.} {\bf B156} (1985) 315.

\bibitem{Metsaev}
R.~R. Metsaev and A.~A. Tseytlin, {\it {Order alpha-prime (Two Loop)
  Equivalence of the String Equations of Motion and the Sigma Model Weyl
  Invariance Conditions: Dependence on the Dilaton and the Antisymmetric
  Tensor}},  {\em Nucl. Phys.} {\bf B293} (1987) 385.

\bibitem{Gross}
D.~J. Gross and J.~H. Sloan, {\it {The Quartic Effective Action for the
  Heterotic String}},  {\em Nucl. Phys.} {\bf B291} (1987) 41.

\bibitem{Sasaki}
N.~Deruelle and M.~Sasaki, {\it {Newton's law on an Einstein 'Gauss-Bonnet'
  brane}},  {\em Prog. Theor. Phys.} {\bf 110} (2003) 441--456,
  [\href{http://xxx.lanl.gov/abs/gr-qc/0306032}{{\tt gr-qc/0306032}}].

\bibitem{Z1}
C.~Charmousis and R.~Zegers, {\it {Matching conditions for a brane of arbitrary
  codimension}},  {\em JHEP} {\bf 08} (2005) 075,
  [\href{http://xxx.lanl.gov/abs/hep-th/0502170}{{\tt hep-th/0502170}}].

\bibitem{Kofinas}
C.~Charmousis, G.~Kofinas, and A.~Papazoglou, {\it {The consistency of
  codimension-2 braneworlds and their cosmology}},  {\em JCAP} {\bf 1001}
  (2010) 022, [\href{http://xxx.lanl.gov/abs/0907.1640}{{\tt
  arXiv:0907.1640}}].

\bibitem{wilt1}
D.~L. Wiltshire, {\it {SPHERICALLY SYMMETRIC SOLUTIONS OF EINSTEIN-MAXWELL
  THEORY WITH A GAUSS-BONNET TERM}},  {\em Phys. Lett.} {\bf B169} (1986) 36.

\bibitem{fax1}
C.~Charmousis and J.-F. Dufaux, {\it {General Gauss-Bonnet brane cosmology}},
  {\em Class. Quant. Grav.} {\bf 19} (2002) 4671--4682,
  [\href{http://xxx.lanl.gov/abs/hep-th/0202107}{{\tt hep-th/0202107}}].

\bibitem{zegers}
R.~Zegers, {\it {Birkhoff's theorem in Lovelock gravity}},  {\em J. Math.
  Phys.} {\bf 46} (2005) 072502,
  [\href{http://xxx.lanl.gov/abs/gr-qc/0505016}{{\tt gr-qc/0505016}}].

\bibitem{deser1}
S.~Deser and J.~Franklin, {\it {Birkhoff for Lovelock redux}},  {\em Class.
  Quant. Grav.} {\bf 22} (2005) L103,
  [\href{http://xxx.lanl.gov/abs/gr-qc/0506014}{{\tt gr-qc/0506014}}].

\bibitem{bowcock}
P.~Bowcock, C.~Charmousis, and R.~Gregory, {\it {General brane cosmologies and
  their global spacetime structure}},  {\em Class. Quant. Grav.} {\bf 17}
  (2000) 4745--4764, [\href{http://xxx.lanl.gov/abs/hep-th/0007177}{{\tt
  hep-th/0007177}}].

\bibitem{Boulware}
D.~G. Boulware and S.~Deser, {\it {String Generated Gravity Models}},  {\em
  Phys. Rev. Lett.} {\bf 55} (1985) 2656.

\bibitem{Cai}
R.-G. Cai, {\it {Gauss-Bonnet black holes in AdS spaces}},  {\em Phys. Rev.}
  {\bf D65} (2002) 084014, [\href{http://xxx.lanl.gov/abs/hep-th/0109133}{{\tt
  hep-th/0109133}}].

\bibitem{papa1}
C.~Charmousis and A.~Papazoglou, {\it {Self-properties of codimension-2
  braneworlds}},  {\em JHEP} {\bf 07} (2008) 062,
  [\href{http://xxx.lanl.gov/abs/0804.2121}{{\tt arXiv:0804.2121}}].

\bibitem{gib}
G.~Gibbons and S.~A. Hartnoll, {\it {A gravitational instability in higher
  dimensions}},  {\em Phys. Rev.} {\bf D66} (2002) 064024,
  [\href{http://xxx.lanl.gov/abs/hep-th/0206202}{{\tt hep-th/0206202}}].

\bibitem{GL}
R.~Gregory and R.~Laflamme, {\it {Black strings and p-branes are unstable}},
  {\em Phys. Rev. Lett.} {\bf 70} (1993) 2837--2840,
  [\href{http://xxx.lanl.gov/abs/hep-th/9301052}{{\tt hep-th/9301052}}].

\bibitem{Dotti}
G.~Dotti and R.~J. Gleiser, {\it {Obstructions on the horizon geometry from
  string theory corrections to Einstein gravity}},  {\em Phys. Lett.} {\bf
  B627} (2005) 174--179, [\href{http://xxx.lanl.gov/abs/hep-th/0508118}{{\tt
  hep-th/0508118}}].

\bibitem{Bogdanos:2009pc}
C.~Bogdanos, C.~Charmousis, B.~Gouteraux, and R.~Zegers, {\it
  {Einstein-Gauss-Bonnet metrics: black holes, black strings and a staticity
  theorem}},  {\em JHEP} {\bf 10} (2009) 037,
  [\href{http://xxx.lanl.gov/abs/0906.4953}{{\tt arXiv:0906.4953}}].

\bibitem{Lanczos}
C.~Lanczos, {\it {A Remarkable property of the Riemann-Christoffel tensor in
  four dimensions}},  {\em Annals Math.} {\bf 39} (1938) 842--850.

\bibitem{Lovelock2}
D.~Lovelock, {\it {The four-dimensionality of space and the einstein tensor}},
  {\em J. Math. Phys.} {\bf 13} (1972) 874--876.

\bibitem{ruthtony}
R.~Gregory and A.~Padilla, {\it {Braneworld instantons}},  {\em Class. Quant.
  Grav.} {\bf 19} (2002) 279--302,
  [\href{http://xxx.lanl.gov/abs/hep-th/0107108}{{\tt hep-th/0107108}}].

\bibitem{Charmousis:2001nq}
C.~Charmousis, {\it {Dilaton spacetimes with a Liouville potential}},  {\em
  Class. Quant. Grav.} {\bf 19} (2002) 83--114,
  [\href{http://xxx.lanl.gov/abs/hep-th/0107126}{{\tt hep-th/0107126}}].

\bibitem{Charmousis:2008ce}
C.~Charmousis and A.~Padilla, {\it {The Instability of Vacua in Gauss-Bonnet
  Gravity}},  {\em JHEP} {\bf 12} (2008) 038,
  [\href{http://xxx.lanl.gov/abs/0807.2864}{{\tt arXiv:0807.2864}}].

\bibitem{ray}
J.~Oliva and S.~Ray, {\it {A new cubic theory of gravity in five dimensions:
  Black hole, Birkhoff's theorem and C-function}},  {\em Class. Quant. Grav.}
  {\bf 27} (2010) 225002, [\href{http://xxx.lanl.gov/abs/1003.4773}{{\tt
  arXiv:1003.4773}}].

\bibitem{Maeda:2010qz}
H.~Maeda, M.~Hassaine, and C.~Martinez, {\it {Magnetic black holes with
  higher-order curvature and gauge corrections in even dimensions}},  {\em
  JHEP} {\bf 08} (2010) 123, [\href{http://xxx.lanl.gov/abs/1006.3604}{{\tt
  arXiv:1006.3604}}].

\bibitem{Emparan:2010ni}
R.~Emparan, S.~Ohashi, and T.~Shiromizu, {\it {No-dipole-hair theorem for
  higher-dimensional static black holes}},  {\em Phys. Rev.} {\bf D82} (2010)
  084032, [\href{http://xxx.lanl.gov/abs/1007.3847}{{\tt arXiv:1007.3847}}].

\bibitem{Agnese:1985xj}
A.~G. Agnese and M.~La~Camera, {\it {GRAVITATION WITHOUT BLACK HOLES}},  {\em
  Phys. Rev.} {\bf D31} (1985) 1280--1286.

\bibitem{Bekenstein:1975ts}
J.~D. Bekenstein, {\it {Black Holes with Scalar Charge}},  {\em Annals Phys.}
  {\bf 91} (1975) 75--82.

\bibitem{Bekenstein:1974sf}
J.~D. Bekenstein, {\it {Exact solutions of Einstein conformal scalar
  equations}},  {\em Ann. Phys.} {\bf 82} (1974) 535--547.

\bibitem{BBM}
K.~B. N.~Bocharova and V.~Melnikov {\em Vestn. Mosk. Univ. Fiz. Astron.} {\bf
  6} (1970) 706.

\bibitem{mad2}
H.~Maeda and N.~Dadhich, {\it {Kaluza-Klein black hole with negatively curved
  extra dimensions in string generated gravity models}},  {\em Phys. Rev.} {\bf
  D74} (2006) 021501, [\href{http://xxx.lanl.gov/abs/hep-th/0605031}{{\tt
  hep-th/0605031}}].

\bibitem{mad}
H.~Maeda and N.~Dadhich, {\it {Matter without matter: Novel Kaluza-Klein
  spacetime in Einstein-Gauss-Bonnet gravity}},  {\em Phys. Rev.} {\bf D75}
  (2007) 044007, [\href{http://xxx.lanl.gov/abs/hep-th/0611188}{{\tt
  hep-th/0611188}}].

\bibitem{cai2}
R.-G. Cai and Q.~Guo, {\it {Gauss-Bonnet black holes in dS spaces}},  {\em
  Phys. Rev.} {\bf D69} (2004) 104025,
  [\href{http://xxx.lanl.gov/abs/hep-th/0311020}{{\tt hep-th/0311020}}].

\end{thebibliography}\endgroup

\bibliographystyle{JHEP}


\end{document}